\documentclass[12pt]{article}
\usepackage{amsmath}
\usepackage{graphicx}
\usepackage{natbib} 
\usepackage{url} 

\newcommand{\blind}{0}

\usepackage{caption}
\usepackage{subcaption}
\usepackage{amssymb}
\DeclareMathOperator*{\argmin}{arg\,min}

\usepackage{times,sectsty}
\subsectionfont{\normalfont\itshape\bfseries}
\subsubsectionfont{\normalfont\itshape}

\begin{document}

\def\spacingset#1{\renewcommand{\baselinestretch}%
{#1}\small\normalsize} \spacingset{1}


\if0\blind
{
  \title{\bf Measuring the Robustness of Predictive Probability for Early Stopping in Experimental Design}
  \author{Daniel Ries 
    \hspace{.2cm}\\
    Statistics and Data Analytics, Sandia National Laboratories\\
    and \\
    Victoria R.C. Sieck \\
    Department of Mathematics and Statistics, Air Force Institute of Technology 
    and \\
    Philip Jones \\
    Aircraft Compatibility Organization, Sandia National Laboratories
    \\
    and \\
    Julie Shaffer \\
    Aircraft Compatibility Organization, Sandia National Laboratories}
  \maketitle
} \fi

\if1\blind
{
  \bigskip
  \bigskip
  \bigskip
  \begin{center}
    {\LARGE\bf Title}
\end{center}
  \medskip
} \fi

\bigskip
\begin{abstract}
Physical experiments in the national security domain are often expensive and time-consuming. Test engineers must certify the compatibility of aircraft and their weapon systems before they can be deployed in the field, but the testing required is time consuming, expensive, and resource limited. Adopting Bayesian adaptive designs are a promising way to borrow from the successes seen in the clinical trials domain. The use of predictive probability (PP) to stop testing early and make faster decisions is particularly appealing given the aforementioned constraints. Given the high-consequence nature of the tests performed in the national security space, a strong understanding of new methods is required before being deployed. Although PP has been thoroughly studied for binary data, there is less work with continuous data, which often in reliability studies interested in certifying the specification limits of components. A simulation study evaluating the robustness of this approach indicate early stopping based on PP is reasonably robust to minor assumption violations, especially when only a few interim analyses are conducted. A post-hoc analysis exploring whether release requirements of a weapon system from an aircraft are within specification with desired reliability resulted in stopping the experiment early and saving 33\% of the experimental runs.
\end{abstract}

\noindent%
\vfill

\newpage
\spacingset{2} 

\section{Introduction}

Design of experiments (DOEx) is a principled way of collecting data to make inferences about a population in a controlled environment. In an engineering setting, this can mean understanding the effects of experimental factors such as temperature or material type on the performance of components, or evaluating the probability of a component operating within specified limits. These experiments are often costly and time consuming, making efficiency paramount to saving time and hardware. There have been many developments in DOEx over the decades to improve efficiency for different scenarios ranging from clinical trials to computer simulations to agricultural experiments. This paper focuses on physical experiments for reliability assessment, meaning the objective is to determine whether a component or part lies within a specification range (commonly referred to as ``spec'') with a certain probability, which is a measure of a component's reliability.  In these situations, testers can be limited in the amount of hardware they have to test and time in a test facility. Bayesian adaptive DOEx is a structured approach to testing developed within the clinical trial field, which typically involves Binomial data. Checking whether a component is within a specification requirement with a certain level of reliability can be done on binary data (e.g. pass/fail), but sometimes a continuous response is a more natural measure. 

One method within Bayesian adaptive DOEx is predictive probability (PP) which evaluates whether the remaining test events need to be seen to adequately assess the performance of the physical system. As with any method, PP is dependent on the data model specification. As such, this paper focuses on understanding the robustness of PP under different data generating models and modeling specifications as a mechanism for Bayesian adaptive DOEx for both situations where the response is continuous or binary.

\subsection{Motivating Application}
\label{sec:motivating}

In order for a fighter aircraft to carry a weapon system, engineers need to first make sure the two can reliably communicate and operate together. One of the ways to make this assessment is to understand the timing transition from aircraft power to weapon power, to ensure the weapon can function as required after it is released from the aircraft. For a weapon system to operate successfully, this timing needs to be within a specification requirement with a certain probability. There is typically a built-up approach to gathering this data, starting with lab umbilical tests, then lab weapon release tests, and ending with aircraft weapon drop flight tests. Lab umbilical pull tests are conducted on the Aircraft Release Simulator (ARS), which was developed to replicate the conditions present when a weapon is separated from a delivery aircraft. It simulates the electrical weapon umbilical cable (power from the aircraft) and actuation pin being pulled free (initiating internal weapon power) at various speeds, temperatures, and pull angles.

\begin{figure}[h]
	\centering
	\includegraphics[width=0.7\textwidth]{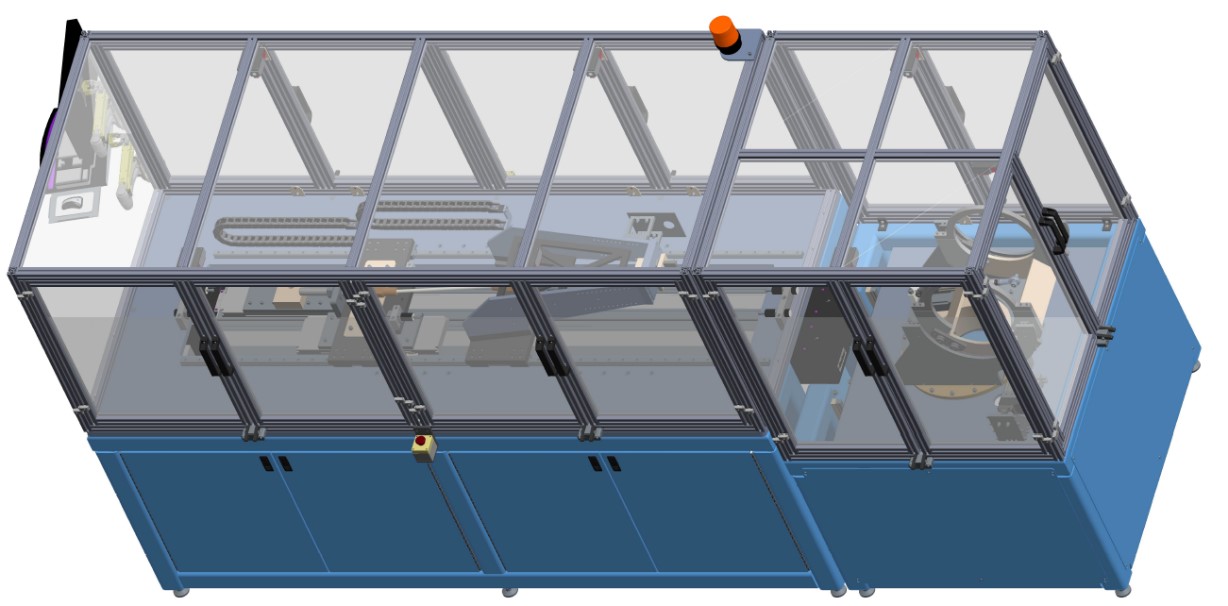}
	\caption{Diagram of pull tester.}
	\label{fig:pulltester-diagram}
\end{figure}

Figure \ref{fig:pulltester-diagram} shows a model of the ARS.
The ARS consists of a section that holds a portion of the weapon case, and a section with a track containing two sleds. The ARS has a data collection system to capture the disconnect timing data for satisfactory separation assessment. High speed cameras are used to examine the qualitative properties of the umbilical and actuation pin release more closely.

\begin{figure}
\centering
\begin{subfigure}{.48\textwidth}
  \centering
  \includegraphics[width=1\textwidth]{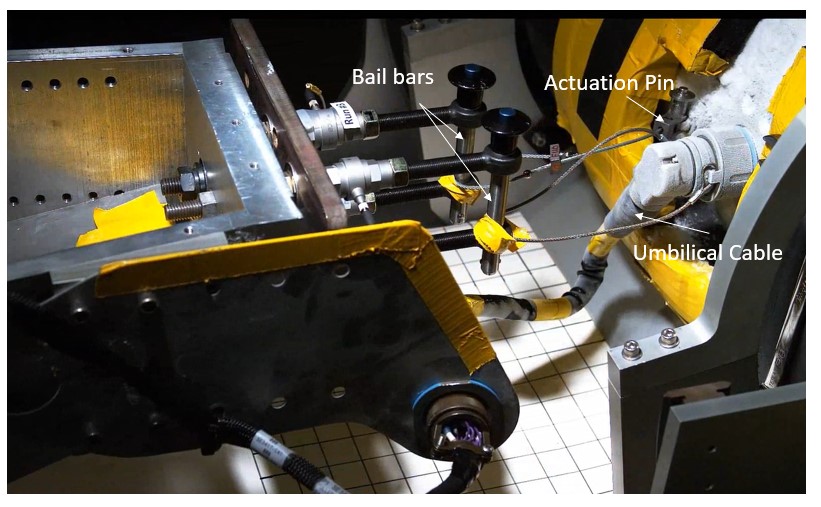}
  \caption{Pull tester aircraft-store interface.}
  \label{fig:pulltester-view1}
\end{subfigure}%
\begin{subfigure}{.48\textwidth}
  \centering
  \includegraphics[width=.9\textwidth]{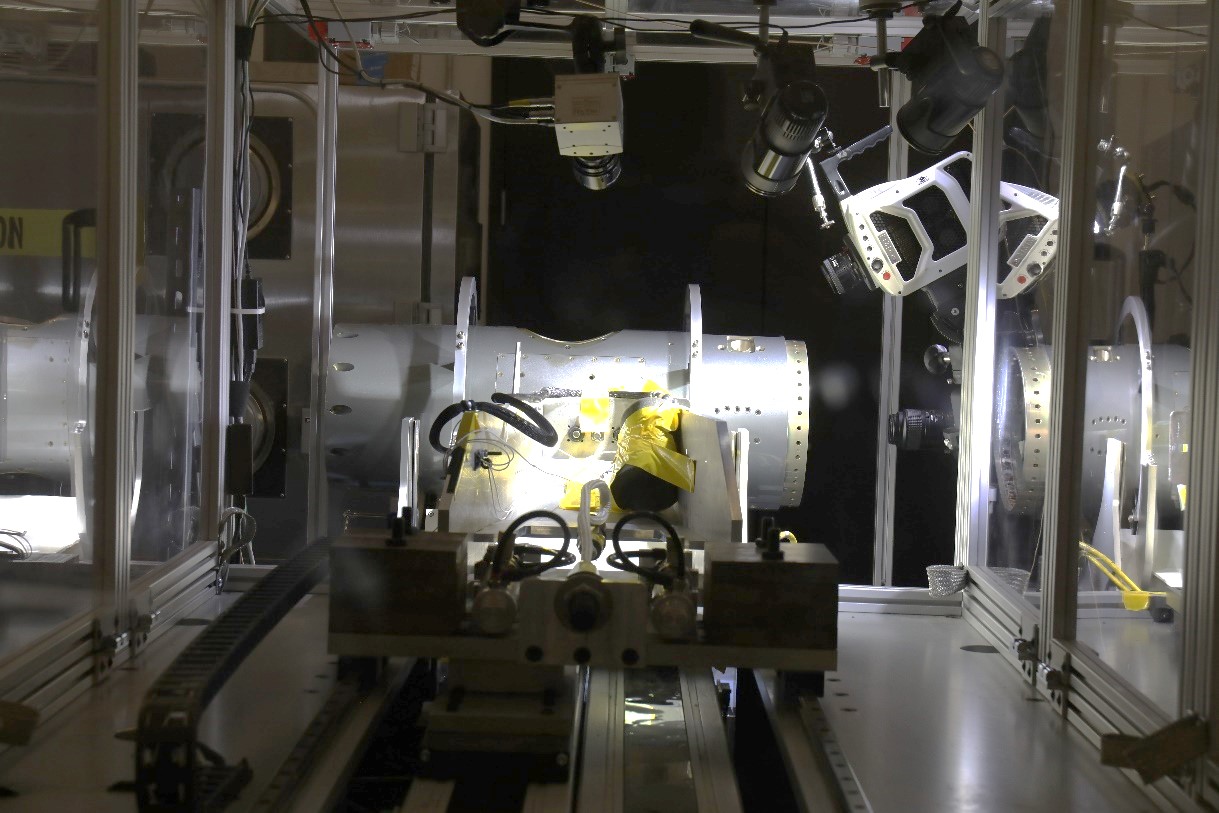}
  \caption{Looking down pull tester track.}
  \label{fig:pulltester-view2}
\end{subfigure}
\caption{View of the pull tester.}
\label{fig:pulltester}
\end{figure}

The first sled is motion controlled and the second sled is attached to the first with a Kevlar rope. Figure \ref{fig:pulltester-view1} shows a view of the pull tester aircraft store interface. When this umbilical cable and actuation pin are pulled from the weapon, the time difference of interest is measured. Figure \ref{fig:pulltester-view2} shows another view of the ARS looking down the tester track.

The second sled is configured to match the geometry of the aircraft being tested. This includes aircraft representative connectors and bail bar(s). The pull tester can be reconfigured to match a given aircraft.  When the pull tester operates, the first, motion-controlled sled moves away from the weapon side at a predetermined speed. This pulls the second sled up to speed very quickly after the rope becomes taut. This allows the second sled to reach velocities seen in actual weapon releases during aircraft drop tests. The second sled pulls away from the weapon case which simulates release of the weapon. 

Historically, a predetermined number of umbilical lab tests for a single weapon/aircraft configuration have been conducted to say with XX\% confidence that for any given umbilical and actuation pin, there is a YY\% probability that it will successfully transition from aircraft to weapon power (XX and YY are omitted for confidentiality reasons). Given that this test is a lower fidelity test to assess weapon satisfactory separation for a high consequence certification evaluation, it was determined to be a suitable candidate to consider adopting Bayesian adaptive DOEx and PP, where it would not only be impactful in reducing test time, but also provide a bridge to gain confidence in the method to adapt it to other higher consequence test activities in the future.

\subsection{Literature Review}

Traditional DOEx ensure optimal statistical properties are obtained using  a fixed sample size, typically using factorial, fractional factorial, or D-optimal designs \citep{morris2010}. These traditional methods have a fixed design where the sample size is typically determined with consideration to statistical power, and the design does not change during testing. However, when testing can be done sequentially it is possible to update the DOEx using data already collected. This is  referred to as adaptive DOEx \citep{berry2011}. Adaptive DOEx can result in changes in many areas of the design including the ability to stop testing early and make a faster conclusion. An adaptive DOEx can determine  the testing can conclude if the data exhibits certain characteristics. 

Bayesian adaptive DOEx is a natural way to do adaptive DOEx since it conditions on what data has been observed. Bayesian adaptive DOEx can consider posterior probabilities or PP, among other quantities, to decide whether to change a design plan. Regardless of the method, the Bayesian approach creates stopping rules based on the probability of possible outcomes. The benefits of this are at least threefold: (i) results are easily interpretable by non-experts, (ii) prior information (e.g. data from previous tests, engineering judgement) can be incorporated into the analysis quantitatively, and (iii)  Bayesian methods condition its findings on all observed data and cohesively account for uncertainties.  This naturally lends itself to adaptive DOEx and updating of a design during an experiment, since at any point during an experiment, the test engineer can stop and reevaluate how to most efficiently complete the experiment. 


Posterior probabilities, computed from the posterior distribution  give the probabilities of characteristics of model parameters given the data. Posterior probabilities  do not consider any unobserved data. Therefore, posterior probabilities are trying to make conclusions about the model parameters given what data has been seen, but it does not take into account the resource limitations of testing. 

The difference between posterior probability and PP is subtle, but important. Posterior probability assesses truth, based on the observed data and any prior information. PP is predicting the outcome of a designed experiment, given the observed data, and integrating over the remaining planned data, therefore it accounts for data still to be collected. As an illustrative example, consider a study interested in knowing whether the probability of the population mean, $\mu$, being greater than 0 is at least 0.9. The DOEx suggests collecting 100 observations sequentially $Z_1,Z_2,...,Z_{100}$, and we will conduct interim analyses at $n_o$=25, 50, and 75. Assume the following data model:

\begin{align*}
Z_i|\mu &\sim \text{Normal}(\mu,1), i=1,2,...,100, \\
p(\mu) &\propto 1\ \text{(Jeffreys prior)}.
\end{align*}

\noindent Let $\bar{Z}_{1:n_o} = \frac{1}{n_o}\sum_{i=1}^{n_o} Z_i$ be the mean of the first $n_0$ observations, and $s^2_{1:n_o} = \frac{1}{n_o}\sum_{i=1}^{n_o} (Z_i-\bar{Z}_{1:n_o})^2$ be the corresponding sample variance. The posterior distribution for $\mu$, after $n_o$ runs, is given by:

\begin{align*}
\mu|Z_1,Z_2,...,Z_{n_o} &\sim \text{Normal} \left(\bar{Z}_{1:n_o},\frac{s^2_{1:n_o}}{n_o} \right).
\end{align*}

\noindent Suppose the following the following were observed: 
\begin{itemize}
\item 
at $n_o=25, \bar{Z}_{1:25}=0.15, s^2_{1:25}=0.99^2$, 
\item 
at $n_o=50, \bar{Z}_{1:50}=0.16, s^2_{1:50}=1.24^2$, 
\item and at $n_o=75, \bar{Z}_{1:75}=0.06,s^2_{1:75}=1.00^2$.
\end{itemize}

The posterior probabilities, $P(\mu > 0|Z_1,...,Z_{n_o})$, computed at $n_o=(25,50,75)$ are $0.78,0.82$, and 0.70, respectively. Although there's a slight downward trend, it is not clear whether seeing the remaining 25 planned observations or not will lead to positive result. The computation of PP will be discussed in Section 2; however, the corresponding PPs, are $0.63, 0.52,$ and  0.09, respectively. Therefore at $n_{75}$, even though the posterior probability at that point is 0.70, the PP is only 0.09. The low PP after seeing 75 observations  could be used as an argument to stop testing early and save 25 tests, something that is difficult to argue with a posterior probability. \citet{saville2014} explains this distinction between posterior probabilities and PPs for Binomial data in a similar way and provides a simple example for that case.

Bayesian adaptive DOEx and PP has a rich history in clinical trials \citep{berry2004,berry2011}. \citet{dmit2006,lee2008,saville2014} each use   PP as the final metric for determining trial futility in simulations and examples. \citet{saville2014} shows how Bayesian PP can be more useful than frequentist p-values or conditional power in adaptive DOEx settings, and even more applicable to stopping rules than Bayesian posterior probabilities.   However, Prior selection can be a concern; \citet{rufibach2016} found that wide, diffuse priors may not reflect that little is known about the parameter of interest. The authors show the effects of many different prior distributions for clinical trial PP which suggests prior sensitivity analyses are important when proposing the use of PP for applications. \citet{broglio2022} compares Bayesian and frequentist adaptive trial designs and finds for their example, the Bayesian approach reached a conclusion faster. Much of the adaptive DOEx literature focuses on clinical trials, and therefore often deals with binary responses.

Predictive probability used on continuous distributions is less common, \citet{geisser1994} showed how PP could be used for interim analyses but resorted to distributional approximations. \citet{zhou2018} developed predictive probability methods that have closed-form solutions for longitudinal data. \citet{lee2008}  developed Bayesian adaptive DOEx for continuous data rather than binary. \citet{liu2018} built off the work of \citet{lee2008} by obtaining a closed form solution for a continuous response. However, their model assumes the population standard deviation is known and only allows for a single interim analysis.  

Although there are examples of adaptive DOEx outside clinical trials, such as in chemistry \citep{misra2017}, materials science \citep{kaneko2021}, engineering reliability \citep{picheny2010}, and manufacturing \citep{pandita2019}, there has been little use of PP for stopping testing early. \citet{sieck2020} introduced Bayesian adaptive DOEx to the quality and reliability engineering fields by taking the concepts built for clinical trials and adapting it to defense applications.  The model used by \citet{sieck2020} was a Normal regression model, providing an example of PP used in physical experiments with continuous responses.

Conditional Power (CP) is a frequentist analog to PP.  \citet{lachin2005} provides an overview of CP with applications to stopping early for futility. \citet{kundu2023} provides a review of CP and PP for continuous and binary random variables. The authors compare some of the properties of CP and PP, and find that CP tends to suggest earlier stopping for futility or efficacy. Because multiple testing can become an issue when using CP, an adjustment procedure is often necessary. The alpha spending function is a frequentist approach commonly used to maintain an overall type I error rate \citep{demets1994}. 

Conversely, there are arguments that Bayesian methods do not require any adjustments because these methods are consistent with the likelihood principle \citep{berry1987}. In practice, there might be cases where this doesn't hold. \citet{ryan2020} explored the performance of PP for type I error rate and power for differing numbers of interim analyses for Binomial data. The authors considered stopping early for futility only, efficacy only, or either. They found type I errors are inflated as the number of interim analyses increase in the stopping for the efficacy only case, but relatively stable in the other two, however, power does suffer. Another question is how to compare PP and CP since they make decisions in different ways: for PP, a decision is made based on posterior probabilities, while CP is based on hypothesis testing. \citet{shi2021} shows the equivalence of p-values and posterior probabilities under one- and two-sided hypothesis tests. Leveraging this approach provides a way of comparing the two approaches.

The remainder of this article is organized as follows, Section 2 introduces the early stopping procedure using PP and CP.  Section 3 introduces the Normal and Binomial reliability models, along with their analogous PP calculations. Section 4 reports on a simulation study evaluating the performance and robustness of both PP and CP, for both the Normal and Binomial reliability models of Section 3. Section 5 applies PP and CP to the application problem introduced in this section. Section 6 summarizes the work in paper, provides ideas for future research and recommendations for the use of Bayesian adaptive DOEx in practice.

\section{Early Stopping during Testing}

Traditional DOEx utilizes a static design, and the models associated with these designs assume the test will be ran until completion before formal statistical analysis begins. However, there are often good reasons to want to be able to stop testing early (e.g. operational testing with limited resources \citep{sieck2020} or clinical trials where ethical concerns come into play \citep{berry2004,palmer2021}). PP and CP can both be used to stop a test early, based on the likelihood of the remaining data  providing a positive or negative result, should the test be run until its conclusion. Unlike posterior probability, PP and CP focus on the the fixed, yet to be seen, resources allocated to the experiment. This section reviews the definitions of PP and CP.

\subsection{Predictive Probability}

Let ${\bf X}=(X_1,X_2,...,X_{n_o})'$ be the data already collected containing $n_o$ observations and ${\bf Y}=(Y_1,Y_2,...,Y_{n_u})'$ be the unobserved data still to be collected containing $n_u$ observations, with the total planned sample size $n = n_o + n_u$. After $n_o$ runs of the test have been completed, the total information consists of the prior information plus the $n_o$ runs.

\subsubsection{General Definition}

In a general experiment, we are interested in determining whether the quantity of interest (QoI), $\phi$, lies above (or below) a threshold, $\phi_0$. Without loss of generality, we consider the case where interest lies in the QoI being greater than the threshold. An evaluation of the QoI with respect to the set of interest is referred to as a \emph{measure}. If $\phi > \phi_0$,  \emph{the measure is met}, otherwise not. We define the probability of the measure being met as $P(\phi > \phi_0)$.  The measure threshold, $\theta_T$, defines a limit that captures the risk a researcher is willing to accept. Therefore in an experimental setting, we say \emph{the measure is met} if $P(\phi > \phi_0) > \theta_T$, and not met otherwise. Within a DOEx context, a frequentist will typically set a type I error rate of $\alpha$, and $\theta_T$ can similarly be thought of as $1-\alpha$ (\citet{shi2021} shows the equivalence under certain scenarios).  

Predictive probability measures the probability that, at some interim point in the test, the measure \emph{would}  have concluded to been met, upon completion of the designed experiment. Formally,

\begin{align}
PP &= P_{Y|X}\left(\mathbf{Y}:P\left(\phi > \phi_0 \middle|\mathbf{X},\mathbf{Y}\right)>\theta_T\right),
\label{eq:pp}
\end{align}

\noindent where $P_{Y|X}$ denotes the predictive distribution of the unobserved data, given the observed data. PP therefore integrates over its uncertainties in parameter estimates, providing a major advantage of Bayesian methods over frequentist methods \citep{berry1993}. 
Given thresholds $\theta_L$ and $\theta_U$, early stopping decisions using PP can be made by:

\begin{align} 
\begin{split}
PP &> \theta_U: \text{stop and declare measure would likely be met,} \label{eq:stoppingDecision1} \\
PP &< \theta_L: \text{stop and declare measure would not likely  be met.}
\end{split}
\end{align}

The thresholds $\theta_L$ and $\theta_U$ are user chosen and account for the level of risk the test engineer is willing to accept. If the threshold $\theta_L$ is set to 0, then the test engineer is only interested in stopping early if the  measure is trending towards being met, this is referred to as efficacy in clinical trials. If the threshold $\theta_U$ is set to 1, then the test engineer is only interested in stopping early if the measure is trending towards not being met, this is referred to as futility in clinical trials. Clinical trials often use PP to check for early signs of drug futility and therefore set $\theta_U$ to 1. They are interested in knowing as soon as possible if the drug is not effective because then they can put patients on another drug or treatment, but do not want to prematurely declare the drug is more effective than the gold standard.

\subsubsection{Predictive Probability for Reliability  }
\label{sec:pp}

The motivation for this study is reducing the number of test units required to show a component is within specifications of a pre-determined reliability requirement.  We consider specification limits  consisting of fixed lower and upper limits, $s_l$ and $s_u$, respectively. The QoI is the component performance within these limits; i.e. the probability the component functions within the limits $s_l$ and $s_u$. We choose $Z$ as a generic random variable for model specification, to avoid confusion when having to distinguish between ``observed'' and ``unobserved'' random variables, $X$ and $Y$. Formally, the QoI is

\begin{align}
\phi &= P\left(s_l<Z<s_u\right).
\label{eq:qoi}
\end{align}

\noindent The QoI in Equation \eqref{eq:qoi} can be placed in the general form in Equation  \eqref{eq:pp}, and PP can be re-expressed as:

\begin{align}
\begin{split}
   PP &= P_{Y|X}\left(\mathbf{Y}:P\left(\phi>\phi_0\middle|\mathbf{X},\mathbf{Y}\right)>\theta_T\right)\\
 &=E_{Y|X}\left(I\left(P\left(\phi>\phi_0\middle|\mathbf{X},\mathbf{Y}\right)>\theta_T\right)\middle|\mathbf{X}\right) \\
&= \int I\left(P\left(\phi>\phi_0\middle|\mathbf{X},\mathbf{Y}\right)>\theta_T\right)  p({\bf Y}|{\bf X}) d{\bf Y} \\
&= \int I\left(\int_{\phi_0}^1 p(\phi|\mathbf{X},\mathbf{Y}) d\phi >\theta_T\right)  p({\bf Y}|{\bf X}) d{\bf Y}.  \label{eq:pp-reliability}
\end{split}
\end{align}

 The inner integral is computing the posterior probability that the measure is met, given the observed data ${\bf X}$ and a  realization of the remaining data, ${\bf Y}$. The indicator function checks whether the measure is met with sufficient probability, as determined by $\theta_T$. This indicator function is then multiplied by the posterior predictive density of the realization ${\bf Y}$, given the observed data. The outer integral marginalizes over all possible realizations of the remaining data ${\bf Y}$. It should be clear PP converges to 1 or 0 as the number of samples observed, $n_o$, approaches the total sample size, $n$. In practice, these integrals can be computed with Monte Carlo integration.

\subsection{Conditional Power}

Conditional power is the probability of achieving (frequentist) statistical significance after all $n$ runs have been completed, given the $n_o$ completed runs, and assumed model parameter values. Common choices for parameter values include the maximum likelihood estimate (MLE), alternative, and null hypothesis values for the parameters. We consider using the MLE here. Statistical significance is defined as observing a test statistic with associated p-value less than a preset level, $\alpha$. The hypotheses are for the problems addressed in this paper are:

\begin{align*}
  H_0: \phi \leq \phi_0, \\
  H_1: \phi > \phi_0.
\end{align*}

Denoting the MLE using observed ${\bf X}$ for $\phi$ as $\hat{\phi}_{\bf X}$, and the test statistic depending on observed and unobserved data, $\xi({\bf X,Y})$, 
the CP according to the defined data model and QoI can be written as:

\begin{align}
\begin{split}
     CP &= P_{Y|\hat{\phi}_{\bf X}}({\bf Y}: P_{\phi_0}(\xi({\bf X,Y})>\Xi_{1-\alpha}^*)) \\
  &= P_{Y|\hat{\phi}_{\bf X}} \left( E_{\phi_0} \left( I \left( \xi({\bf X,Y})>\Xi_{1-\alpha}^*  \right) \right) \right)\\
    &= \int \left( \int I \left( \xi({\bf X,Y})>\Xi_{1-\alpha}^*  \right) d{\bf X} \right)  p({\bf Y}|\hat{\phi}_{\bf X}) d{\bf Y}, \label{eq:cp-reliability}
\end{split}
\end{align}

\noindent where $\Xi_{1-\alpha}^*$ is the $1-\alpha$ quantile of $\Xi$, the distribution of the test statistic $\xi({\bf X,Y})$, $P_{Y|\hat{\phi}_{\bf X}}$ denotes the sampling distribution of the unobserved data ${\bf Y}$ under a likelihood with parameters taking values equal to $\hat{\phi}_{\bf X}$, and $P_{\phi_0}$ and $E_{\phi_0}$ denote the probability and expectation under the null hypothesis, respectively. \citet{saville2014} gives equations for CP and PP in the binary case.


\section{Models for Reliability}

The application presented in Section 1 focused on a reliability problem where the response was a continuous variable. However, reliability problems also often deal with binary responses much like clinical trials. This section introduces both a Normal reliability model for continuous response data and a Binomial reliability model for binary data. Define ${\bf Z} = (Z_1,Z_2,...,Z_n)'$ as random variables; 
 $Z_i \in\mathbb{R}$ for the continuous case and $Z_i \in \{0,1\}$ for the Binary case.

\subsection{Normal Model}
\label{sec:normal}

The commonly used 2-parameter Normal model is considered.  The distribution for $Z_i$ can be written as:

\begin{align}
Z_i|\mu,\sigma^2 &\sim \text{Normal}\left(\mu,\sigma^2\right),i=1,...,n.
\label{eq:likelihood}
\end{align}

\noindent A conjugate prior for this model is:

\begin{align*}
 \sigma^2 &\sim \text{Inverse-Gamma}(a,b), \\
 \mu|\sigma^2 &\sim \text{Normal} \left( m,\frac{\sigma^2}{\nu} \right).
\end{align*}

\noindent The hyperparameters have easy interpretations: $m$ is the prior mean based on $\nu$ observations, and the variance $\sigma^2$ is based on $2a$ observations with a sum of squares equal to $2b$. The resulting posterior distribution is:

\begin{align}
\begin{split}
\sigma^2|\mathbf{Z} &\sim \text{Inverse-Gamma}\left(a',b'\right), \\
\mu|\sigma^2,\mathbf{Z} &\sim \text{Normal}\left(m',\frac{\sigma^2}{\nu' }\right), \\
f\left(\mu,\sigma^2|\mathbf{Z}\right) &=\ f(\mu|\sigma^2,\mathbf{Z})\ f\left(\sigma^2|\mathbf{Z}\right), \label{eq:posterior}
\end{split}
\end{align}

\noindent where $f(\cdot|\cdot)$ denotes a conditional density function, and:

\begin{align*}
    a' &= a+\frac{n}{2}, \\
    b' &= b+\frac{1}{2}\sum_{i=1}^{n\ }(Z_i-{\bar{Z})}^2 + \frac{n\nu}{\nu+n}\frac{(\bar{Z}-m)^2}{2}, \\
    \nu' &= \nu + n, \\
    m' &= \frac{\nu m + n \bar{Z}}{\nu'}.
\end{align*}

\noindent The posterior predictive distribution for a future observation  $Z^\ast$, is also available in closed form:

\begin{align}
Z^\ast|\mathbf{Z} \sim t_{2\left (\alpha + \frac{n}{2} \right)} \left(m',\frac{b'(\nu'+1)}{\nu'a'} \right). \label{eq:posterior-predictive}
\end{align}

\noindent The predictive distribution can be used to sample the unobserved data required to induce a distribution on the QoI required to calculate PP in Section 2.1.2.  From the PP definition in Equation \eqref{eq:pp-reliability}, $p({\bf Y}|{\bf X})$ is Monte Carlo (MC) sampled from the distribution in Equation \eqref{eq:posterior-predictive},  and $p(\phi|{\bf X},{\bf Y})$ is MC sampled by taking draws of $\mu$ and $\sigma^2$ from Equation \eqref{eq:posterior}, and computing $\phi = \Phi\left( \frac{s_u-\mu}{\sigma} \right) - \Phi\left( \frac{s_l-\mu}{\sigma} \right)$, where $\Phi(\cdot)$ is the standard Normal cumulative density function. From the CP definition in Equation \eqref{eq:cp-reliability}, $p({\bf Y}|\hat{\phi}_{\bf X})$ is MC sampled from the density in Equation \eqref{eq:likelihood}, where $\hat{\phi}_{\bf X} = \Phi\left( \frac{s_u-\hat{\mu}_{{\bf X}}}{\hat{\sigma}_{{\bf X}}} \right) - \Phi\left( \frac{s_l-\hat{\mu}_{{\bf X}}}{\hat{\sigma}_{{\bf X}}} \right)$ using MLEs of $\mu$ and $\sigma^2$, $\hat{\mu}_{{\bf X}}$ and  $\hat{\sigma}_{{\bf X}}$, respectively, estimated using ${\bf X}$. The test statistic $\xi({\bf X,Y}) = \Phi\left( \frac{s_u-\hat{\mu}_{{\bf X,Y}}}{\hat{\sigma}_{{\bf X,Y}}} \right) - \Phi\left( \frac{s_l-\hat{\mu}_{{\bf X,Y}}}{\hat{\sigma}_{{\bf X,Y}}} \right)$ where $\hat{\mu}_{{\bf X,Y}}$ and  $\hat{\sigma}_{{\bf X,Y}}$ are MLEs using  ${\bf X}$, and the ${\bf Y}$ sampled from $p({\bf Y}|\hat{\phi}_{\bf X})$. In this setup, it is apparent that $\xi({\bf X,Y}) = \hat{\phi}_{{\bf X,Y}}$, the MLE for $\phi$ using ${\bf X}$ and sampled ${\bf Y}$. The distribution $\Xi$ is generated by MC simulation.

\subsection{Binomial Model}

For the case where response data is binary, we consider a Binomial model.  The distribution of $Z_i$ can be written as:

\begin{align*}
 Z_i|p &\sim \text{Bernoulli}(p),\ i=1,2,...,n,
\label{eq:likelihood2}
\end{align*}

\noindent A conjugate prior for this model is:

\begin{align*}
 p &\sim \text{Beta}(\alpha,\beta).
\end{align*}

\noindent The hyperparameters have easy interpretations: $\alpha$ represents the number of ``prior'' successes, and $\beta$ represents the number of ``prior'' failures  out of $\alpha + \beta$ ``prior'' trials. 

The posterior distribution is:

\begin{align*}
    p|{\bf Z} &\sim Beta(\alpha',\beta'), \\
    \alpha' &= \alpha + \sum_{i=1}^n Z_i, \\
    \beta' &= \beta + n - \sum_{i=1}^n Z_i.
\end{align*}

\noindent The posterior predictive distribution for for a future observation $Z^\ast$, is also available in closed form:

\begin{align*}
    Z^\ast|\mathbf{Z} &\sim \text{Beta-Binomial}(\alpha',\beta').
\end{align*}

\noindent As with the predictive distribution for the Normal data model, the predictive distribution can be used to obtain the unseen observations required to induce a distribution on the QoI required to calculate PP in Section 2.1.2. Computation of PP for the Binomial model case does not require MC sampling and is explained in \citet{saville2014}.

\section{Simulation Studies}

Both PP and CP are dependent upon their model specification in order to integrate over unobserved data. A natural question is: how robust are PP and CP to their model specification? Practitioners need to specify a data model under incomplete knowledge in order to use PP or CP, so it is important to understand the relative importance of their assumptions. This section describes a simulation study to assess their robustness of early stopping decisions. Both the Normal and Binomial models are considered.

\begin{figure}[h]
			\centering
			\includegraphics[width=0.7\linewidth]{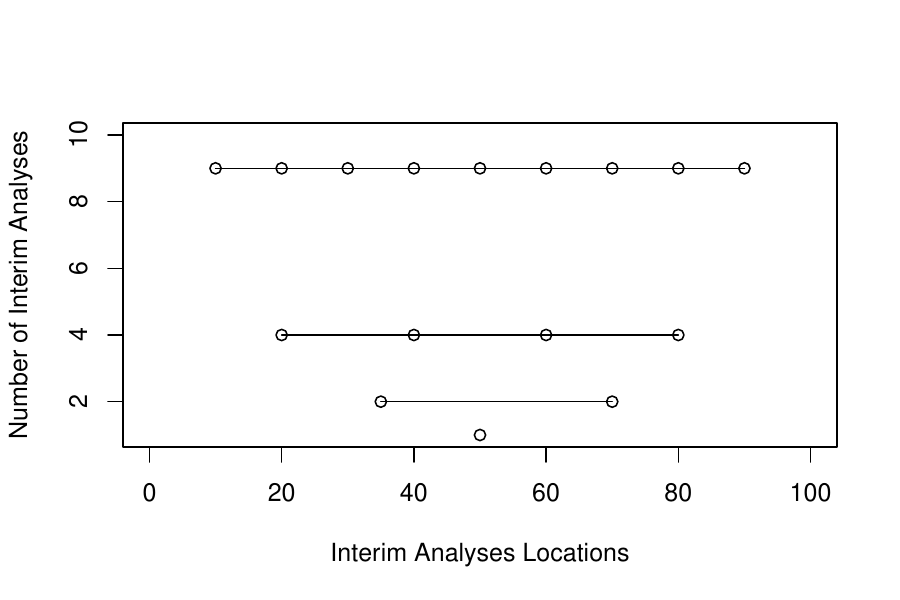}
			\caption{Timings of interim analyses for a 100-run design. Total number of interim analyses for a design is given by y-axis.}
			\label{fig:interim-analysis-plot}
		\end{figure}

To evaluate whether a specification limit is met, we may be interested in stopping a test early for efficacy only, futility only, or  for either efficacy or futility. In this study, we consider designs that are originally planned to conduct 100 tests, and consider designs with zero, one, two, four, and nine interim analyses.  Figure \ref{fig:interim-analysis-plot} gives a visual representation of when these interim analyses are conducted. This same general design will be used for both the Normal model and Binomial model simulations. 
		
We evaluate model performance and compare PP and CP using standard metrics. Type I error rates come from scenarios where the measure is erroneously concluded to be met, and power comes from scenarios where measure is correctly concluded to be met. These metrics are based on the first instance where $PP <\theta_L$ (for futility), $PP >\theta_U$ (for efficacy), or $PP <\theta_L \cup PP >\theta_U$ (for either), for each simulated data set. The same procedure holds for CP. We set the posterior threshold $\theta_T=0.95$ and frequentist significance level $\alpha=0.05$. Early stopping decisions for either are made with $\theta_L=0.05$, and $\theta_U=0.95$. For futility only, $\theta_U=1$, for efficacy only, $\theta_L=0$. For early stopping, CP follows the same procedure as PP given in Equation \eqref{eq:stoppingDecision1}. We correct for multiple testing with CP using an alpha spending function. This is implemented using the \texttt{getDesignGroupSequential} function in the \emph{rpact} R package \citep{rpact}.

\subsection{Normal Model Case}

 We use the Normal model from Section \ref{sec:normal}, and the QoI, $\phi$, is the probability of being within specification, $P\left(s_l<Z<s_u\right)$ and $\phi_0$ is the threshold value that must be obtained for the measure to be met. 

\subsubsection{Normal Data Generating Mechanism(s)}
\label{sec:dgm}

The data generating mechanism (DGM) for the simulated data for the Normal model varies the distributional family so there are several distributional shapes. The model in Equation \eqref{eq:likelihood} assumes Normally distributed data, so here we explore violations to that assumption. The bounds for the specification, as in Equation \eqref{eq:qoi}, are set to $s_l=2$ and $s_u=5$. The probability of the measure being met is  $P(\phi > 0.8)$, and type I error rates come from scenarios where the true $\phi=0.8$, and power comes from scenarios where the true $\phi=0.9$ and the conclusion that the measure was met.

The distribution families considered are Normal($\mu,\sigma^2$), Laplace($\mu,\delta$), Uniform($\gamma_L,\gamma_U$), and two mean-shifted Gamma($\alpha_k,\beta_k$), $k=L,H$. For the mean-shifted Gamma, two parameterizations are considered, one with relatively high skewness ($k=H$) and one with relatively low skewness ($k=L$).  
Ensuring each of the distributions has the correct $\phi$ value requires optimizing for distributional parameters. As close as possible, expected values are set to be equal to 3.5 and scale parameters are adjusted accordingly to accommodate the respective $\phi$ for the DGM. In case of shifted-Gamma distribution, the scale parameter is set to a constant for both the low- and high-skew versions and the entire distribution is shifted to the right by 2. Parameters for the 10 DGMs (5 distributions $\times$ 2 values of $\phi$), with specified properties, are obtained by:

\begin{align*}
 \hat{\sigma}_m^2 &=\ \argmin_{\sigma^2}{\left(\int_{2}^{5}\frac{1}{\sqrt{\left(2\pi\sigma^2\right)}}e^{-\frac{1}{\sigma^2}\left(w-\mu\right)^2}dw-\phi_m\right)^2},m=1,2, \\
  \hat{\delta}_m &=\ \argmin_{\delta}{\left(\int_{2}^{5}\frac{1}{2\delta}e^{-\frac{|w-\lambda|}{\delta}} dw-\phi_m\right)^2},m=1,2, \\
 {\hat{\alpha}}_{L,m} &=\ \argmin_{\alpha}{\left(\int_{2}^{5}\frac{1}{\Gamma\left(\alpha\right)\beta_L^\alpha}w^{\alpha-1}e^{-\frac{w}{\beta_H}}dw-\phi_m\right)^2},m=1,2,\\
  {\hat{\alpha}}_{H,m} &=\ \argmin_{\alpha}{\left(\int_{2}^{5}\frac{1}{\Gamma\left(\alpha\right)\beta_H^\alpha}w^{\alpha-1}e^{-\frac{w}{\beta_L}}dw-\phi_m\right)^2}, m=1,2,\\
 {\hat{\gamma}}_{L,m}\ &= \argmin_{\gamma_L}{\left(\int_{2}^{5}\frac{1}{{\gamma_U-\gamma}_L}dw-\phi_m\right)^2},m=1,2,
\end{align*}

\noindent where $\mu=3.5,\lambda=3.5,\beta_L=4, \beta_H=2, \gamma_U=3.5\times 2-\gamma_L, \phi_1=0.8, \phi_2=0.9$. A figure displaying the probability density functions of all 10 DGMs is provided in the supplemental material. The distribution of the test statistic $\xi({\bf X,Y})$ is obtained by simulation, and a histogram of this simulated distribution is included in the supplemental material.

One thousand data sets are created from each of the 10 DGMs. At every $n_o$ where PP and CP are computed, 1000 possible realizations of the $n_u$ unobserved data are generated. Because PP depends on prior specification, we present results using a relatively benign and uninformative prior ($m=3.5,\nu=1,a=1,b=1$) which results in a prior probability of the measure being met, $P(\phi > 0.8)=0.29$. A full sensitivity analysis for the prior selection is included in the supplemental material.

\subsubsection{Results}

\begin{figure}[h]
\centering
\includegraphics[height=.4\textheight]{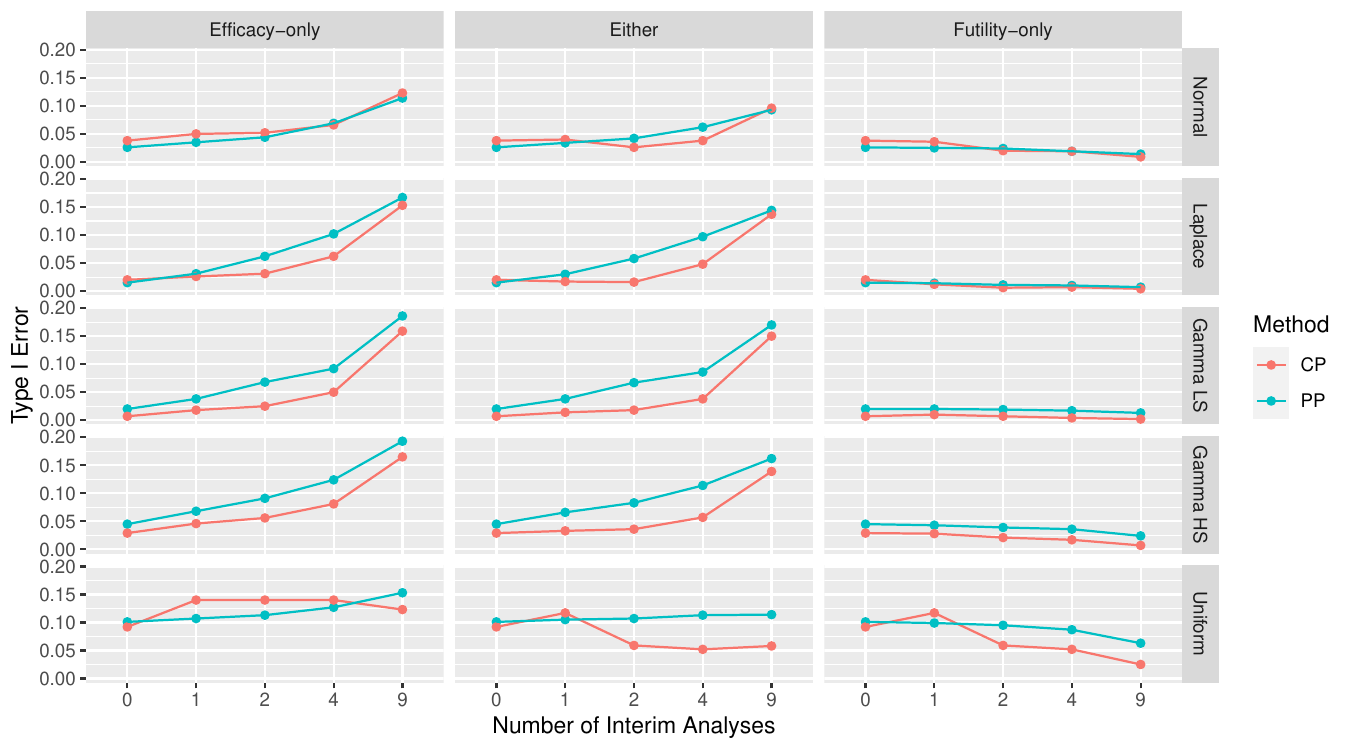}
\caption{Type I error for PP and CP, with stopping rules in the columns and DGMs in the rows. }
\label{fig:type1}
\end{figure}

Figure \ref{fig:type1} shows type I error across different stopping decision rules and DGMs for PP and CP. Overall, CP tends to have lower type I error rates than PP, although CP corrects for multiple comparisons with an alpha spending function. Type I error rates for non-Normal distributions tend to be low for a small number of interim analyses while their rates increase faster than the Normal DGM. This holds for both the stopping for efficacy only and stopping for either cases. The Uniform DGM doesn't follow this trend as nicely and tends to see much higher type I error rates even at low numbers of interim analyses. When considering stopping early only for futility, neither PP nor CP have type I error issues, except with a Uniform distribution. This suggests a degree of robustness for PP and CP when there is only interest in stopping a test early to claim futility.

\begin{figure}[h]
\centering
\includegraphics[height=.4\textheight]{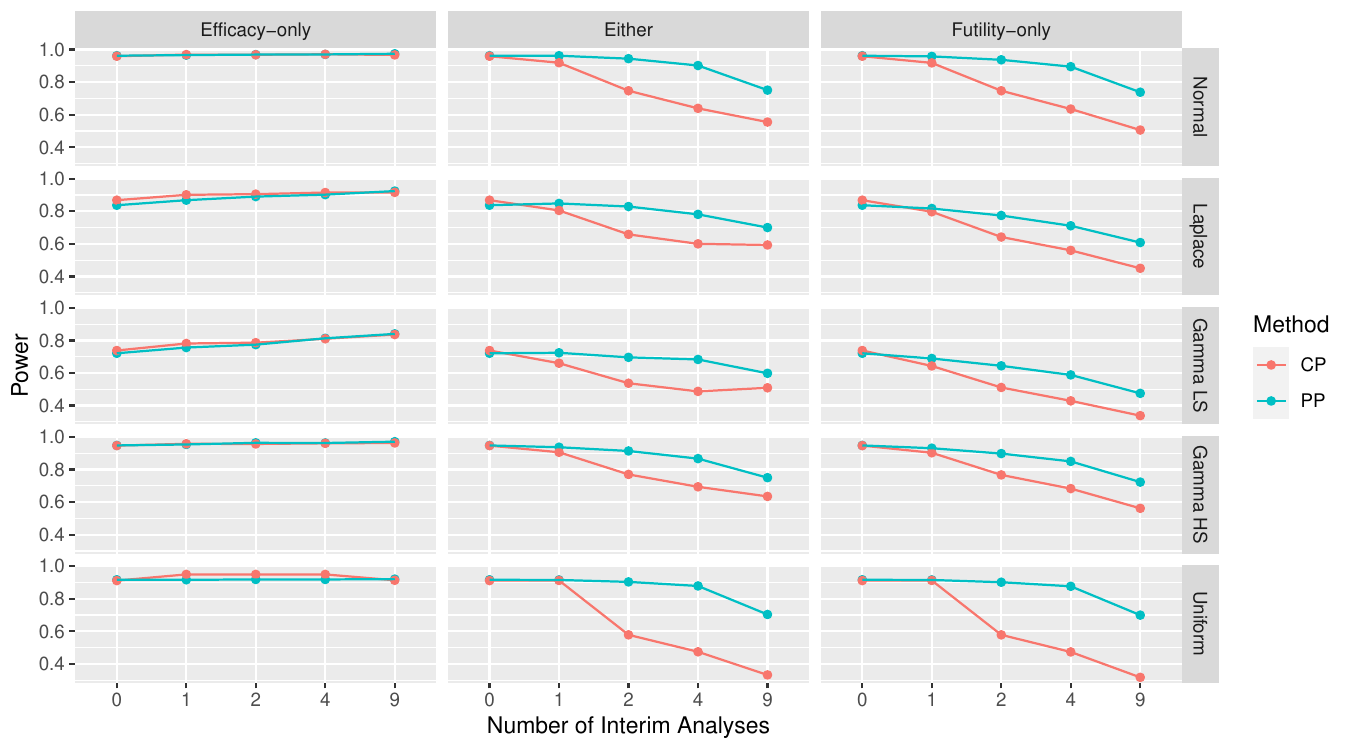}
\caption{Power for PP and CP, with stopping rules in the columns and DGMs in the rows. }
\label{fig:power}
\end{figure}

Figure \ref{fig:power} shows power for different stopping decision rules and DGMs for PP and CP. Both PP and CP have high and comparable power for the efficacy-only case. In other scenarios, PP has much larger power, especially as the number of interim analyses increases. For all DGM, PP has power greater than 0.7 no matter the stopping rule or the number of interim analyses, showing robustness to model specification. This is especially important in the stopping for futility-only case, since this gives the model many opportunities to stop testing early and make an incorrect decision. When considering two or more interim analyses, PP power is at least 10\% higher, and in many cases  more than 15\% higher.

\begin{figure}[h]
\centering
\includegraphics[width=\textwidth]{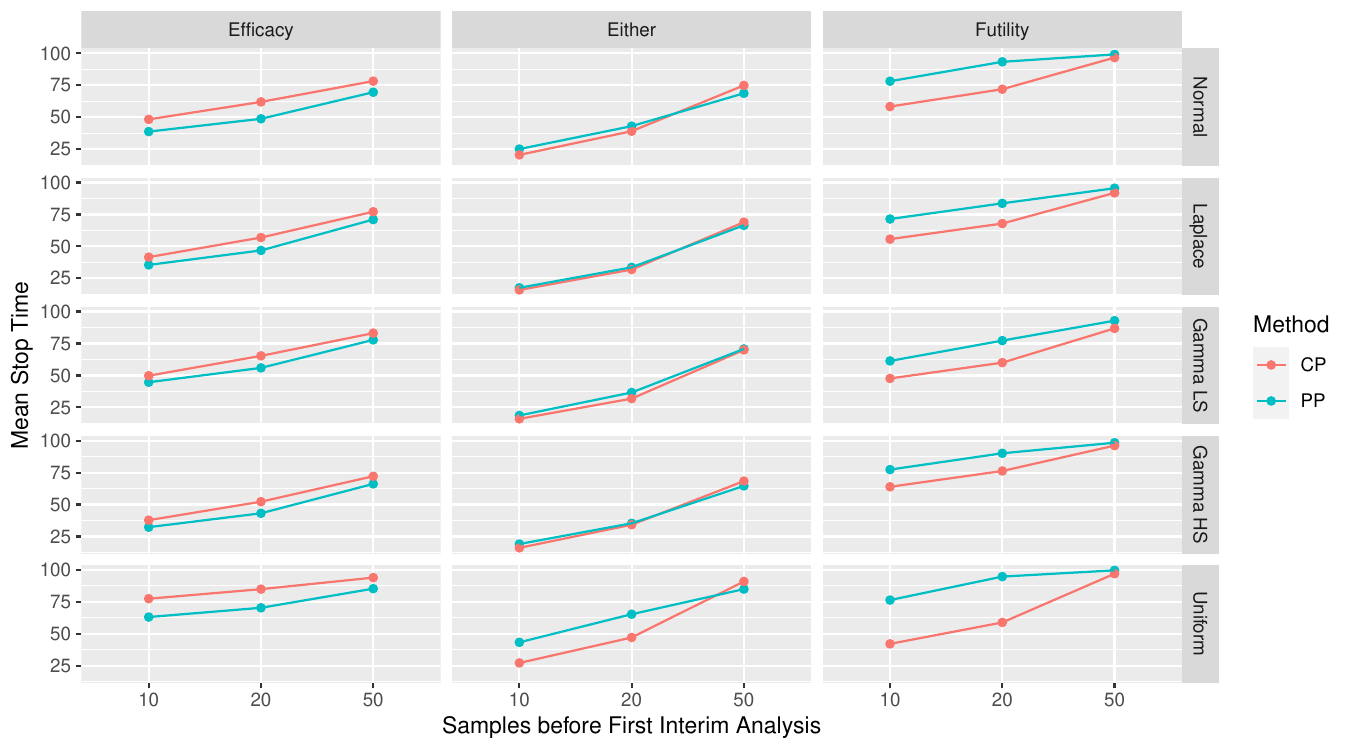}
\caption{Mean stopping times based on number of samples before first interim analysis for $\phi=0.9$ case. }
\label{fig:stopping_times}
\end{figure}

Figure \ref{fig:stopping_times} shows the mean stopping times for all DGMs for both PP and CP for $\phi=0.9$ case. This figure quantifies how quickly a first stopping decision is made by PP and CP under different scenarios. Predictive probability tends to make decisions quicker than CP in the efficacy-only case, whereas CP tends to make decisions quicker in the futility-only case, and both are similar in the either case. It is interesting that both CP and PP will stop earlier for the Gamma DGM than it will for the Normal DGM.

\begin{figure}[h]
\centering
\includegraphics[width=\textwidth]{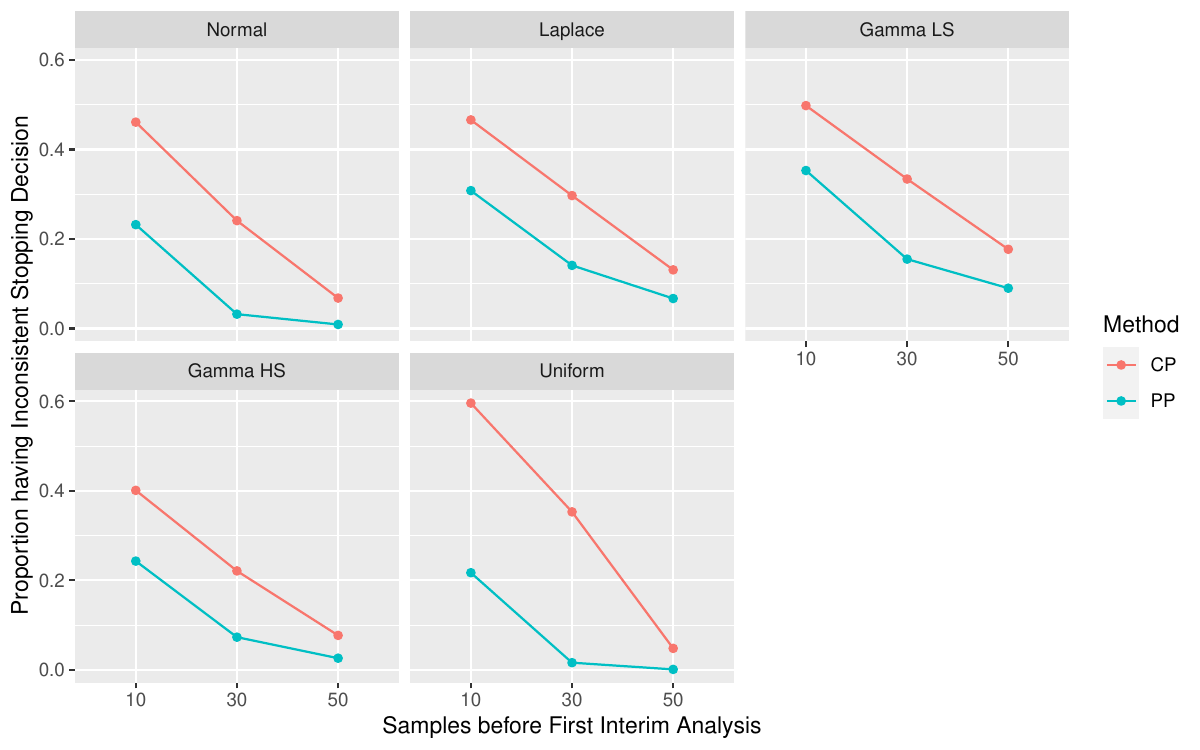}
\caption{Proportion of inconsistent early stopping decisions. This quantifies the proportion of times the first early stopping decision \emph{would have been reversed, i.e. stopping for the opposite reason,} had more data been observed before the first stopping decision was originally made. }
\label{fig:inconsistent}
\end{figure}

Because this is a simulation study, we can explore if the approaches would ``flip'' their stopping decision later in the design and produce a different decision after seeing more of the data. This allows us to understand the stability of conclusions, which are an important argument when presenting such an approach decision makers who are not experts in statistics or DOEx. Figure \ref{fig:inconsistent} shows the proportion of inconsistent early stopping decisions for all DGMs for both PP and CP. Conditional power tends to change its mind at a much higher rate than PP, indicating it is less stable. We believe this makes sense since CP isn't integrating over its uncertainty in its parameter estimates, and as such, it not  considering possible outcomes of remaining experiments in the same way PP does. From a practitioner's perspective, this type of analysis can also be used to help determine the number of experimental runs to observe before doing a first interim analysis, since we want a model's conclusions to be stable. In a real data analysis, we can't check how consistent or inconsistent a model will be, so we could make this decision based on simulated examples. Here, observing about 30\% of the data before the first interim analyses appears to be a good trade-off between efficiency and making erroneous mistakes.

\subsection{Binomial Model Case}

Consider an adaptive testing approach where the response is binary, as is common among destructive testing where the response is pass/fail or go/no-go. The QoI, $\phi$, is the probability of success (same as Binomial parameter $p$) and $\phi_0$ is the threshold value that must be obtained for the measure to be met.

\subsubsection{Binomial Data Generating Mechanism(s)}

For the Binomial case, we generate 100 datasets independently each from a Binomial$(100, 0.6)$ and from a Binomial$(100, 0.75)$. Type I errors for this scenario come from decisions to stop testing early and claim the measure is met when the data was simulated from Binomial$(100, 0.6)$. Similarly, power comes from from decisions to stop testing early and claim the measure is met when the data was simulated from Binomial$(100, 0.75)$. The question of interest for the Binomial case is determining whether $\phi > 0.60$ (i.e. $\phi_0 = 0.60$). As above, we present results using a relatively benign and uninformative prior ($\alpha=1,\beta=1$) which results in a prior probability of the measure being met, $P(\phi > 0.8)=0.2$. As with the Normal model, we include a sensitivity analysis for the Binomial case in the supplemental material.

\subsubsection{Results}

\label{sec:dgm2}

\begin{figure}[h]
	\centering
	\includegraphics[width=\linewidth]{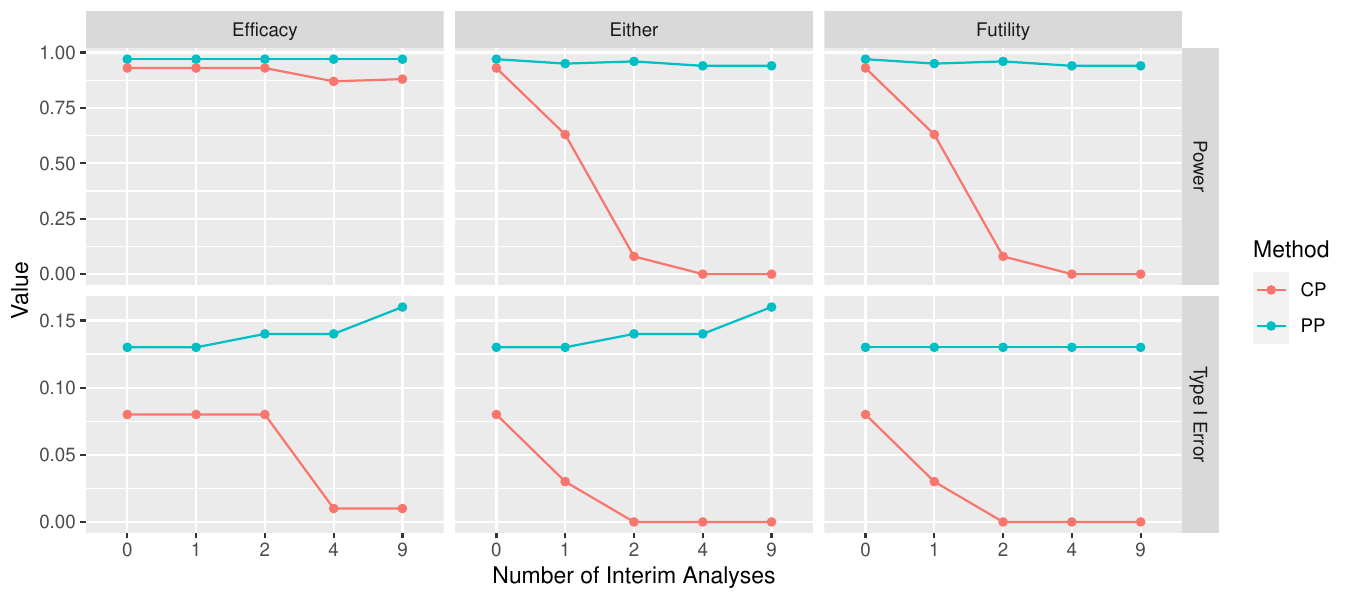}
	\caption{Power and Type I error for PP and CP, with stopping rules in the columns.}
	\label{fig:type-i-and-powerboth}
\end{figure}

Figure \ref{fig:type-i-and-powerboth} shows the Type I error and power results for PP and CP on the Binomial case. Overall, PP has higher power, and its power tends to remain fairly constant as the number of interim analyses increase. This is in contrast to CP, whose power significantly degrades as the number of interim analyses increases, for the futility-only and either cases. Conversely, PP has higher Type I error rates than CP,  although the rates don't increase much as the number of interim analyses increase. From a practical perspective, a multiple testing adjustment may be required for PP if Type I error rates need to be controlled.

While simulations suggest that CP has better type I error rates for this type of set-up, the simulations further suggest that CP can have very low power when stopping for futility, or stopping for either futility or efficacy, is allowed. Examining these results further, it was found that the driving factor for the behavior of power seen in Figure \ref{fig:type-i-and-powerboth} was largely driven by the choice of the first time interim analysis was accomplished. While the number of interim analyses did play a role in the decrease of power, it was minimal compared to the point at which interim analysis was first conducted. For instance, Figure \ref{fig:type-i-and-powerboth} demonstrates power using the Binomial(100, 0.75) data sets. The first of the 9 interim analyses started at 10 observations seen (and continuing every 10 observations until either a decision was made or all 100 observations had been seen); had we considered CP when the first of 9 interim analyses was 91 (and continuing every observation until either a decision was made or all 100 observations had been seen), the power would be 0.93. Alternatively, if we considered two interim analyses at 50 and 70 (instead of the established 35 and 70), power was comparable to one interim analysis at 50 observations. Explorations suggest that evaluating CP before to seeing at least 50 observations is significantly detrimental to the power of the method. Therefore, careful consideration should be given to picking the first point of interim analysis when using CP to stop for futility or either futility or efficacy under a construct as proposed in this example. 

\begin{figure}[h]
	\centering
	\includegraphics[width=\linewidth]{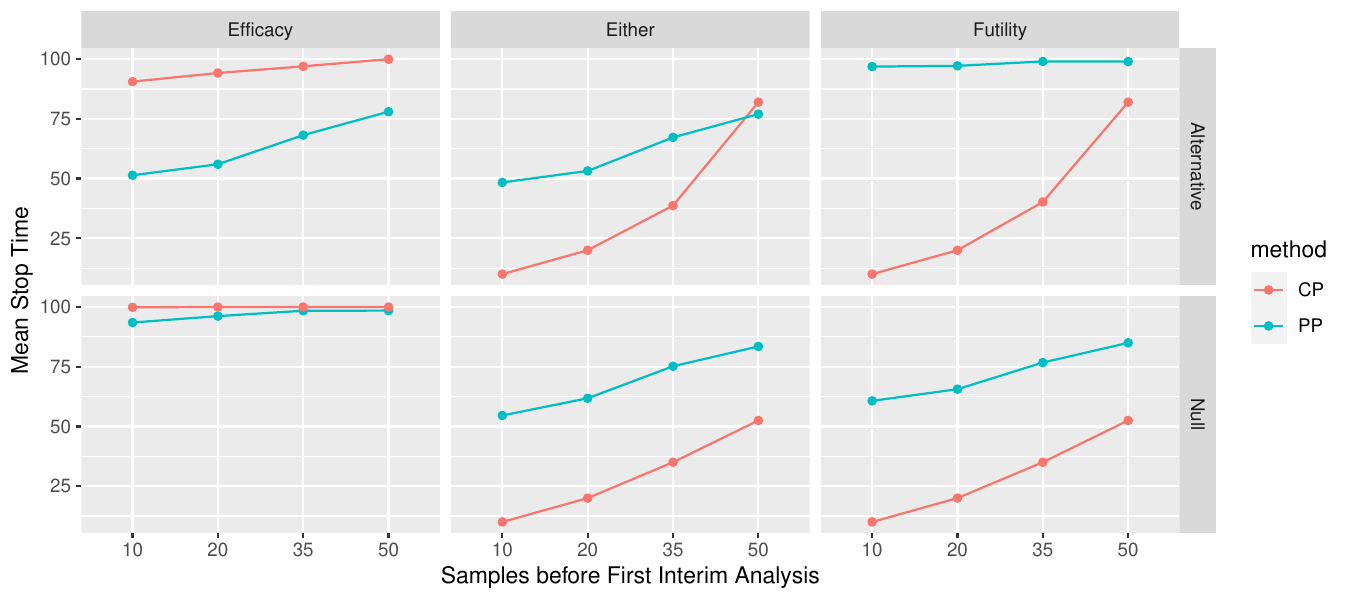}
	\caption{Average Stopping times for CP and PP under null ($p=0.6$) and alternative ($p=0.75$) hypotheses.}
	\label{fig:mean-stoppingone-case}
\end{figure}

To further understand this behavior, Figure \ref{fig:mean-stoppingone-case} demonstrates the average point in the test in which PP and CP stop, under both the null ($p=0.6$) and alternative ($p=0.75$) hypotheses. These plots demonstrate that CP is, making a decision much earlier in the test than PP for the futility-only and either case. This early decision, specifically when related to stopping a test for futility, is overly conservative when using alpha spending, impacting the power of the test. Alternatively, for efficacy-only PP has faster stopping times paired with higher power.

\section{Application}

Recall from Section \ref{sec:motivating}, we are interested in assessing whether pull times meet specification limits for a given reliability level. These pull times simulate the difference in time between when the umbilical cable and actuation pin are disconnected from the weapon upon ejection. The variable we want to determine the reliability of is the difference in time between the when the umbilical cable detaches and the pull out switch assembly actuation pin detaches. Therefore, the measurable in this experiment is a scalar value. The original design called for 180 pulls on the ARS in order to achieve the desired reliability level. The test engineers said the experiment would need to run for a minimum of 30 pulls, so the first look occurs at $n_o$=30. For confidentiality purposes, we cannot present the raw timing data, and specifications given here are notional. Our measure is $P(s_l < T < s_u) > \phi_0$ with $s_l=-3$, $s_u=3$, $\phi_0=0.95$. Following the simulation study, we used a benign prior of $\mu_0=0, \nu=1, a=1, b=1$, although we considered a variety of priors, and include a sensitivity analysis  in the supplemental material.

Figure \ref{fig:app-results} shows a post-hoc analysis conducted in the same order as the original data collection. The black line shows the PP re-calculated after every 10 pulls, the red line shows CP for the same situation. The red dashed line is the upper threshold $\theta_U$=0.95. This threshold was first met at 120 pulls, which corresponds to a savings of 33\% compared to the originally designed experiment. Comparatively, using CP would have stopped at 90 pulls, with a savings of 50\%. However, the simulation study showed PP to be a more conservative approach than CP, making it more applicable to this high-consequence problem. Had PP been used in this situation for early stopping, significant resources in time and hardware could have been saved.

\begin{figure}[h]
		\centering
\includegraphics[width=.6\textwidth]{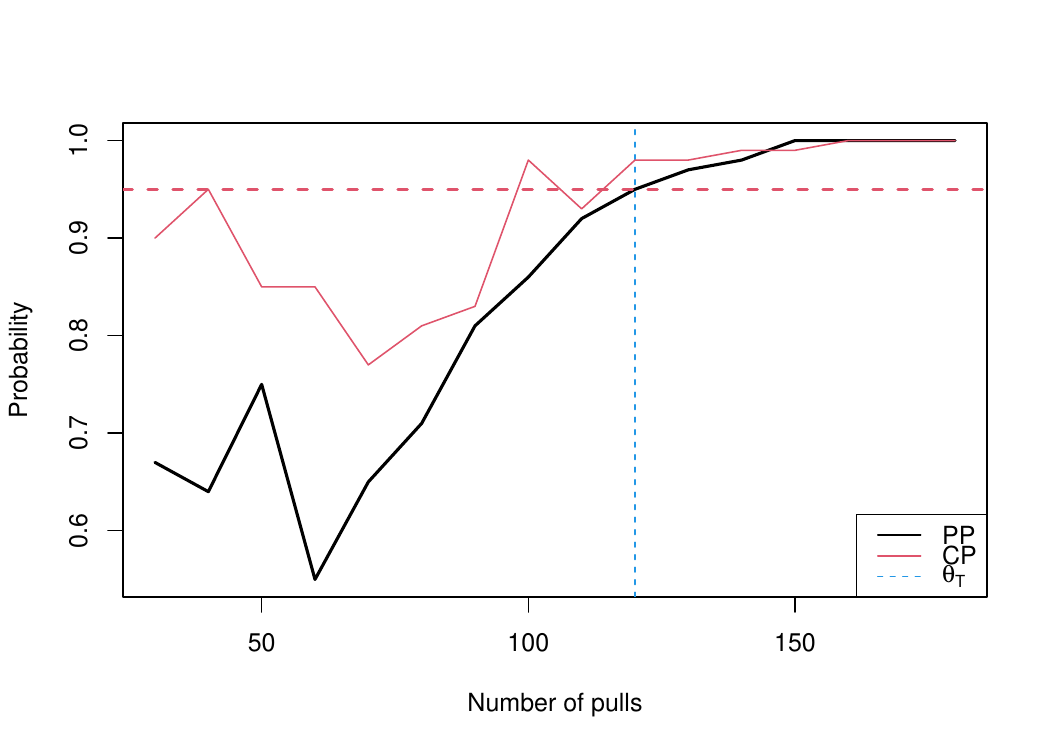}
\caption{PP (black solid line) and CP (red solid line) for pull timing application. Red horizontal dashed line represents $\theta_T$, the stopping threshold.}
\label{fig:app-results}
\end{figure}

\begin{figure}[h]
		\centering
\includegraphics[width=.6\textwidth]{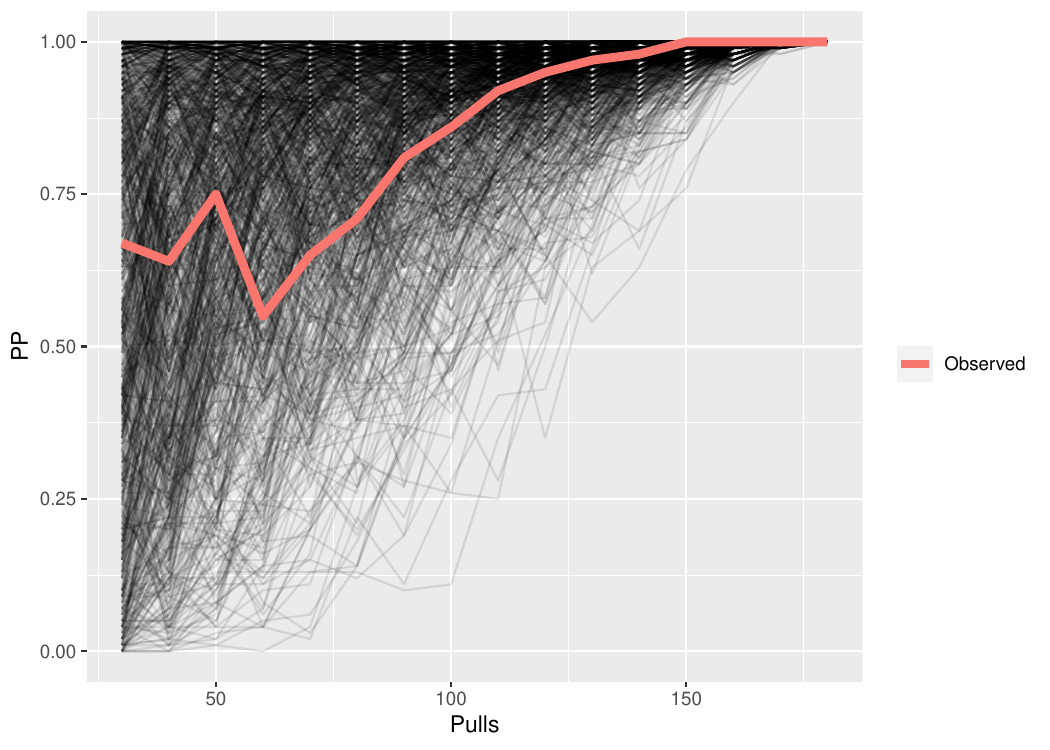}
\caption{One thousand permutations of the pull timing data, with the observed order in bolded red. }
\label{fig:app-perm}
\end{figure}

Because this was a post-hoc analysis and the original design was randomized, it could be argued we observed the data in a favorable order. To alleviate this issue, we permuted the order of the experiment 1000 times and calculated PP for each permutation over the length of the experiment. Figure \ref{fig:app-perm} shows the resulting PP curves, with the observed PP in red. In 5.5\% of scenarios, $PP < \theta_L = 0.05$, meaning the incorrect stopping decision (stop testing for futility) would have been made.

These results show the significant benefit PP could have provided. The results were consistent across different experimental orderings, providing a degree of confidence in the method which could be explained to decision makers. This was considering a relatively benign prior, expert prior information could have been easily included into this analysis which could result in even further efficiency gains.

\section{Discussion}

This paper takes a closer look at Bayesian adaptive DOEx and the use of predictive probabilities as a means to stop experimental testing early, with a particular emphasis on physical, engineering-focused experiments. Stopping tests early due to futility or efficacy can result in cost-, time-, and resource-savings. Predictive probabilities are the Bayesian solution to this problem since they measure the probability of concluding the experimental goal, or measure, would be attained if an experiment were to be completed. While there can be significant savings in time and resources by stopping testing early and making a decision, there can also be significant costs if this decision is not well informed and incorrect. Predictive probabilities  are dependent on the model, which could have significant implications for stopping early since it requires integrating over unobserved data. To understand when practitioners could use predictive probabilities, we conducted a robustness study to understand how predictive probabilities behave when modeling assumptions break down. 

The first model considered was a Normal model when the response data is continuous.  Practitioners who use PP for early stopping and want Type I error rate controlled should consider a multiple testing adjustment given these results, at least when looking for efficacy-only or either efficacy or futility. Early stopping decisions using PP appeared relatively robust to distribution changes, except for the Uniform DGM. PP has robust Type I error rates for early stopping for futility-only, without any multiple testing adjustment. When considering power, PP is robust to distribution changes in the efficacy-only scenario. There are some degradations, especially at a high number of interim analyses for stopping for either futility or efficacy, or futility only. Based on these simulations, the Normal model is relatively robust to varying distributions, no matter the end goal, as long as a test engineer accepts only considering one or two interim analyses. This could result in significant savings, even in the case of one interim analysis, stopping for efficacy-only could save between 10-25\% of the tests. Simulation results for the Binomial model for pass/fail data describe a similar story. PP tended to have higher than nominal Type I error rates, but very high power. 

When PP was applied to the application of pull testing on aircraft release mechanisms, the results showed potential for significant savings. Had PP been used during testing, 33\% fewer pulls than originally scheduled could have been done. With the simulation study to show the robustness of PP to deviation from the Normal model, test engineers could feel more comfortable in  stopping a test early to declare their measure is met or not. 

Generally speaking, CP has a better type I error rate, while PP has better power. Therefore, trade offs should be considered with respect to type I error rate and power when selecting between PP and CP. Furthermore, CP only makes quicker decisions about system performance in the futility only case—although CP may be overly conservative when using alpha spending to correct for multiple interim analyses, leading to erring on the side of failing to meet requirements. 

Finally, the simulation study also suggested the lack of stability with CP. In the reliability case where the data was assumed to be normal, CP had a higher rate of “flipping” it’s decision, ultimately leading to the suggestion that at least 30\% of observations be seen before employing CP. Furthermore, the binomial reliability case demonstrated that CP is very sensitive to, not the number of interim analyses, but rather the location of those interim analyses. This case suggested that at least 50\% of observations be seen before employing CP. Overall, given the potential pitfalls of CP, it is recommended that PP be used instead of CP.

Future work should continue to understand the behavior of early stopping with PP  for different DGMs than the ones considered here, and for different models than a simple Normal model. Including experimental factors in a linear and a non-linear way are promising steps forward, as well as other assumption modifications such as non-constant variance. Having a comprehensive understanding of  PP based on different DGMs will help practitioners build confidence in the method because they will know what to expect when the data looks a certain way, and how to proceed in such scenarios. At this stage, significant statistics expertise is needed to run Bayesian adaptive DOEx, and all early stopping decisions should be made jointly between the statisticians and subject matter experts to properly account for the risks and benefits of stopping testing early. As more studies such as this and applications emerge using Bayesian adaptive DOEx, the more comfortable the community will become with this powerful approach. 

\section{Acknowledgements}
The authors thank 
J. Gabriel Huerta from Statistical Sciences organization at Sandia National Laboratories for his helpful comments and edits. Sandia National Laboratories is a multimission laboratory managed and operated by National Technology \& Engineering Solutions of Sandia, LLC, a wholly owned subsidiary of Honeywell International Inc., for the U.S. Department of Energy's National Nuclear Security Administration under contract DE-NA0003525. This paper describes objective technical results and analysis. Any subjective views or opinions that might be expressed in the paper do not necessarily represent the views of the U.S. Department of Energy or the United States Government.  SAND2023-10055O.

\section*{Data Availability Statement}

The data used for the application is proprietary information and cannot be shared. The simulated data is available upon request.

\bibliography{bibliography}
\bibliographystyle{Perfect}

\clearpage

\section*{Supplemental Material}

The supplemental material contains additional information on the simulated data for the Normal  and Binomial models' simulation study and prior sensitivity analyses, and the prior sensitivity analysis of the application.

\subsection*{Simulated Data for Normal Model}

\begin{figure}[h]
\centering
\includegraphics[width=\textwidth]{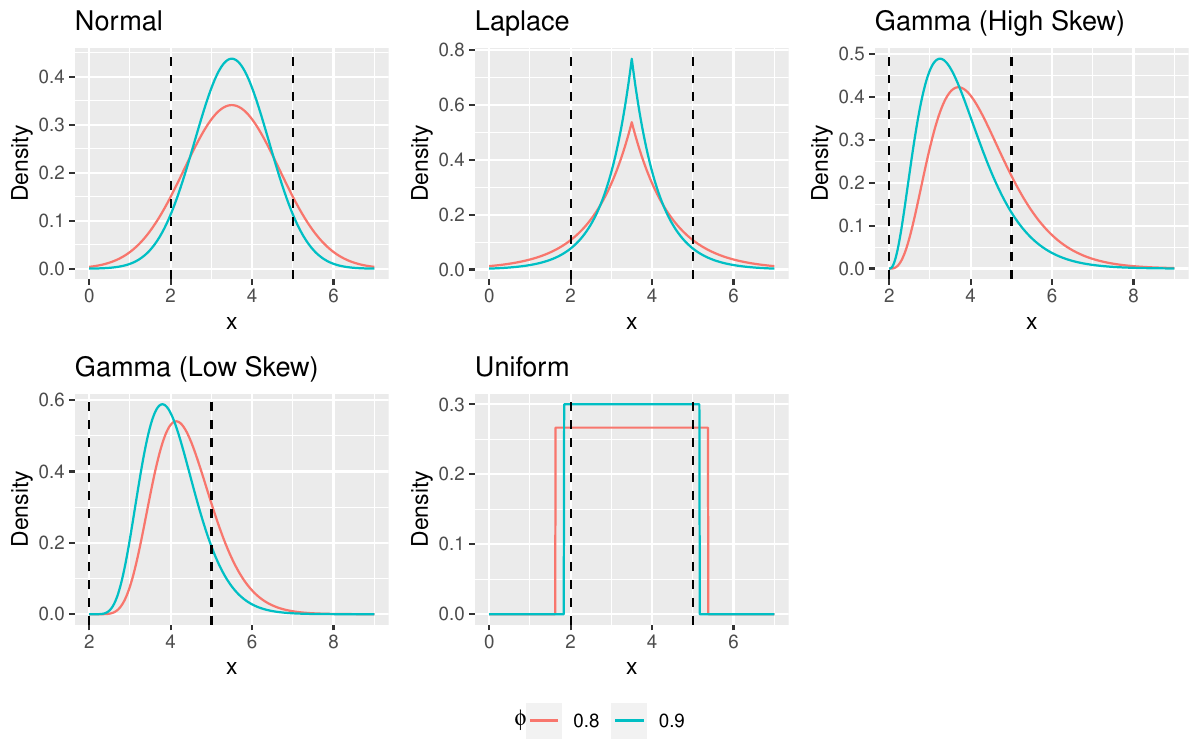}
\caption{Probability density functions of all 10 DGMs. Lines are colored according to true $\phi$. Vertical dashed lines indicate the reliability bounds, $a$ and $b$, for the reliability QoI. }
\label{fig:dgm}
\end{figure}

\begin{figure}[h]
\centering
\includegraphics[width=0.8\textwidth]{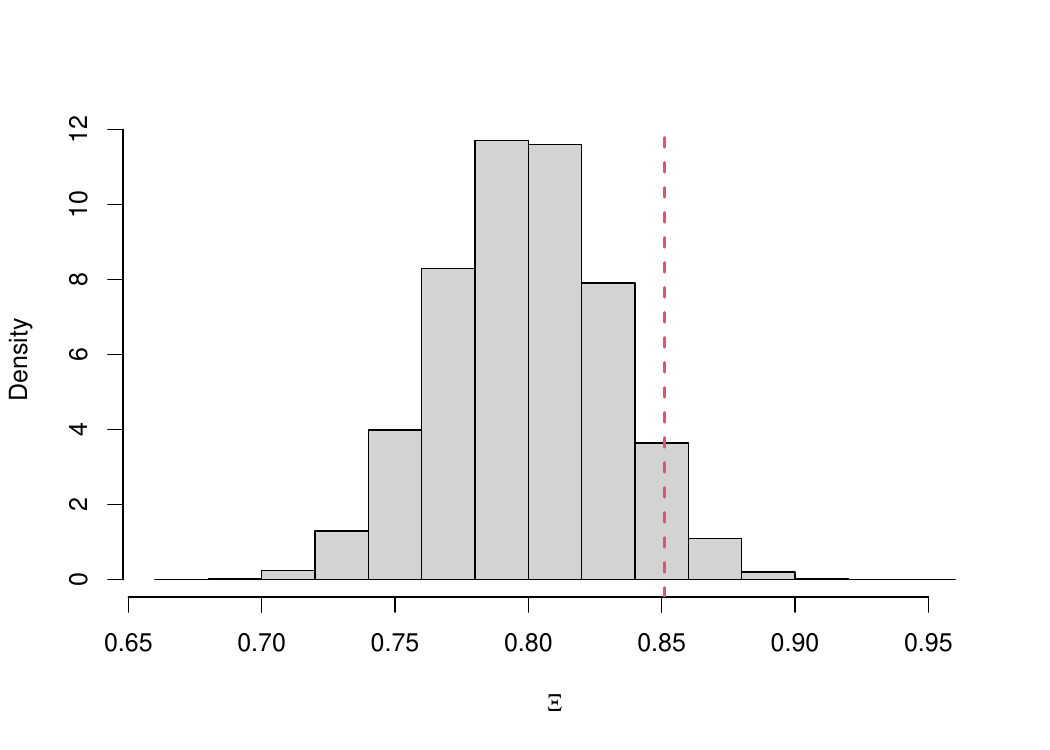}
\caption{Simulated distributions of the test statistic, $\xi({\bf X,Y})$ with $\Xi_{0.95}^*$ denoted by red dashed line.}
\label{fig:test_statistic}
\end{figure}

Figure \ref{fig:dgm} shows the probability density functions (PDF) of the DGMs for the simulation study for the Normal model. The value of $\phi$ represents the QoI, and is the probability between the dashed lines in each PDF. Figure \ref{fig:test_statistic} shows the simulated distribution of the test statistic, $\xi({\bf X,Y})$, used  in the simulation study for the Normal model case. The red dashed line indicates the 95$th$ percentile. 

\subsection*{Prior Sensitivity Analysis for Normal Model Case}		

\begin{figure}[h]
	\centering
	\includegraphics[height=0.9\textheight]{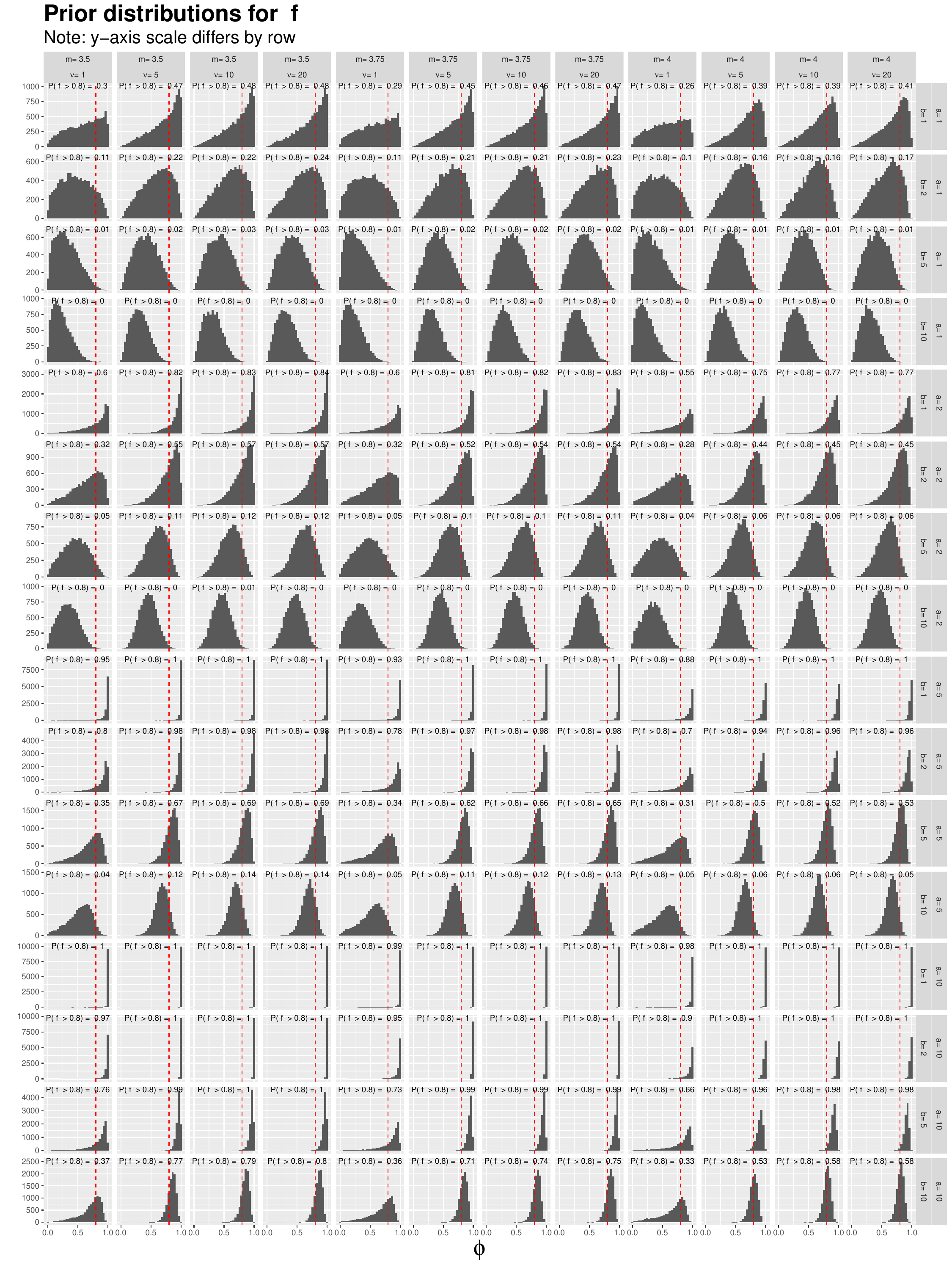}
	\caption{Prior distributions for $\phi$ considered in sensitivity analysis of simulation study. Prior probability of measure being met, $P(\phi > 0.8)$ is shown for each prior. The red dashed line indicates $\phi=0.8$.}
	\label{fig:priors}
\end{figure}

In order to understand the influence of the selected prior in the main paper, we performed a prior sensitivity analysis for the Normal model. Figure \ref{fig:priors} shows the different prior distributions for $\phi$ considered. The values of the hyperparameters $m, \nu, a,b $ are given in the facets. The resulting prior probability of the measure being met, $P(\phi > 0.8)$ is given in text on each plot. The red dashed line indicates $\phi=0.8$. We considered a wide variety of shapes for priors, some which are relatively flat, some are strong, and some are relatively weak. 

\begin{figure}[h]
	\centering
	\includegraphics[height=0.9\textheight]{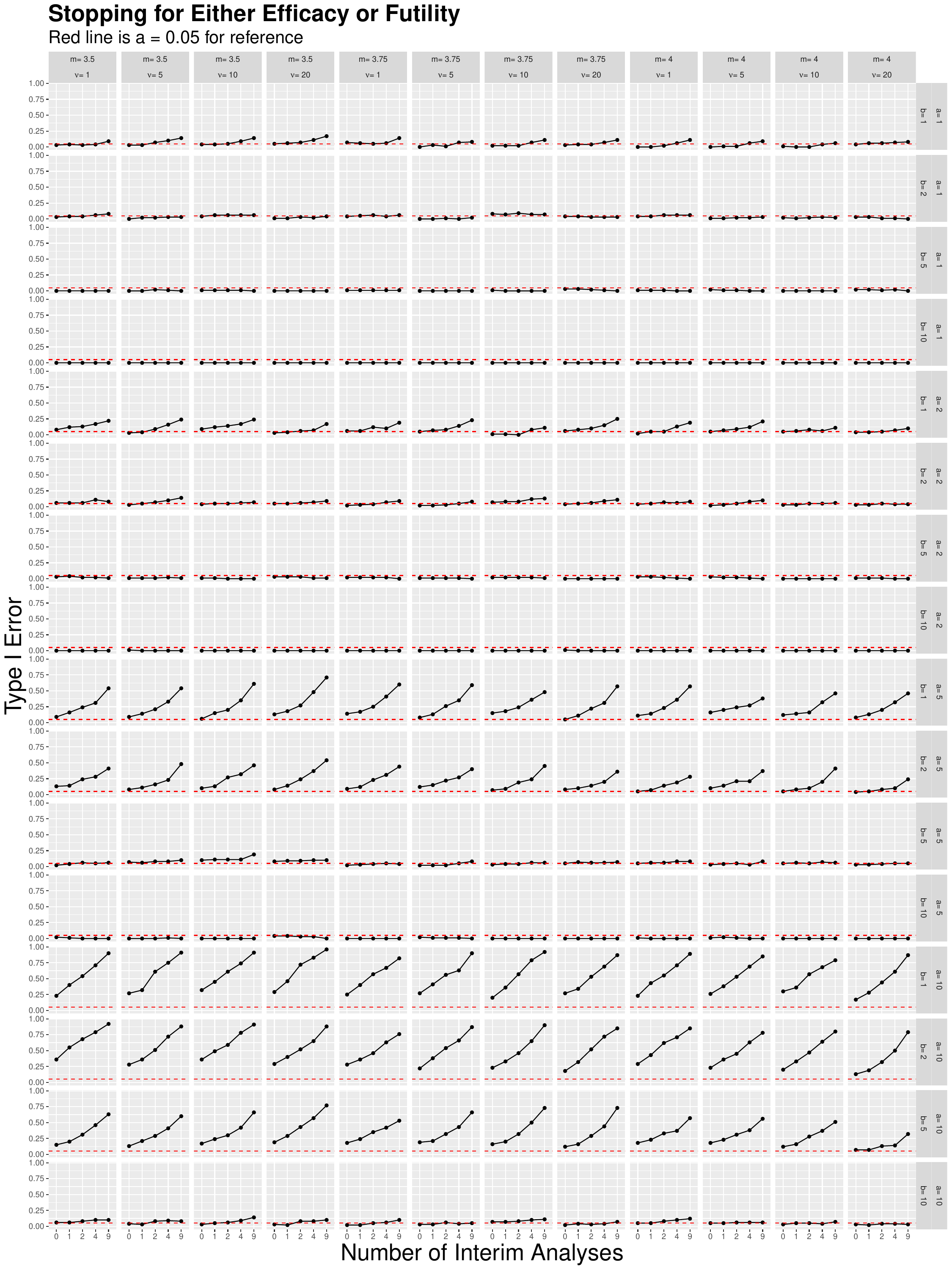}
	\caption{Type I errors when stopping for either futility or efficacy for simulation study prior sensitivity analysis.}
	\label{fig:sm-either-type1}
\end{figure}

\begin{figure}[h]
	\centering
	\includegraphics[height=0.9\textheight]{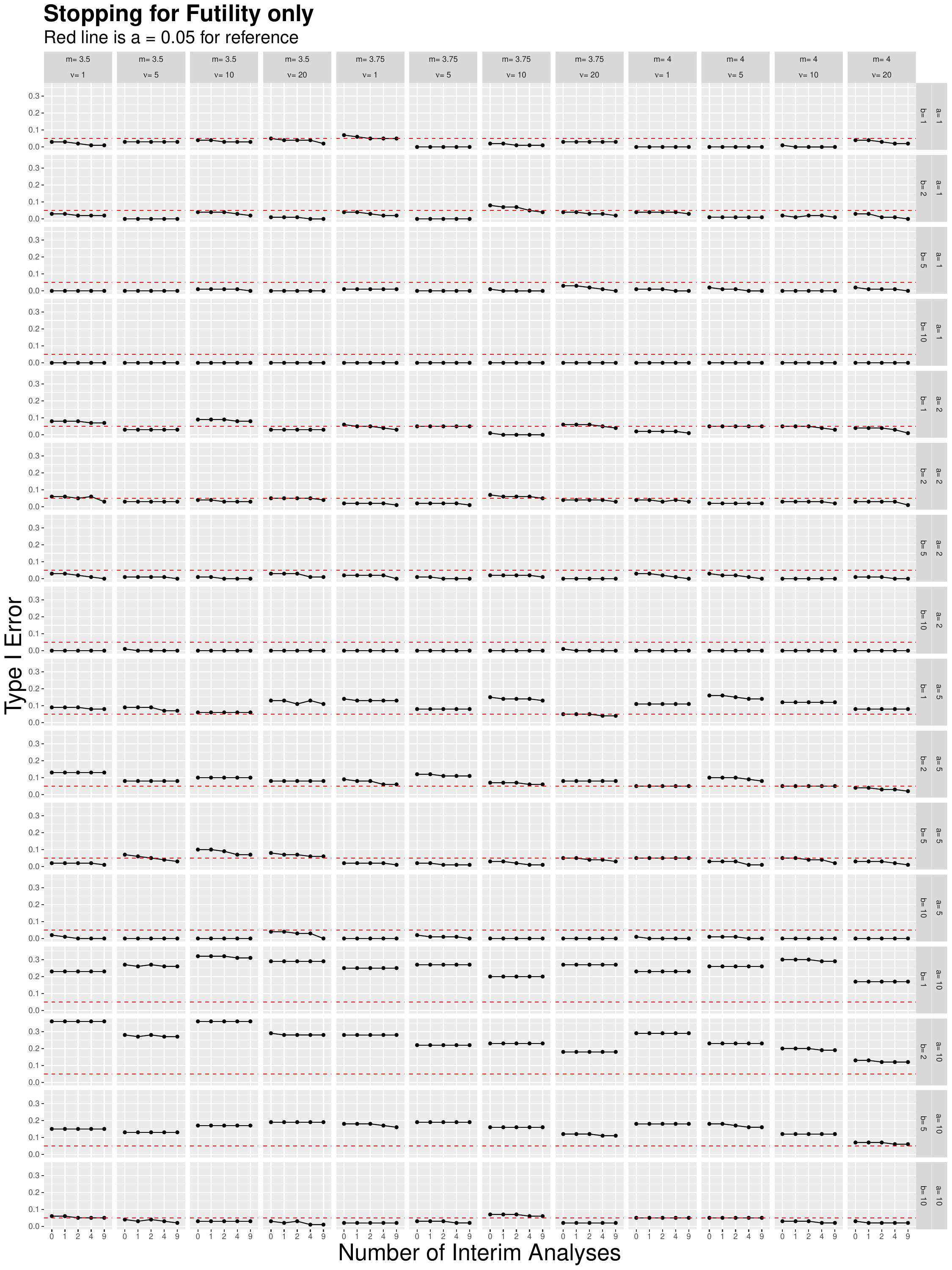}
	\caption{Type I errors when stopping for futility for simulation study prior sensitivity analysis.}
	\label{fig:sm-futility-type1}
\end{figure}

\begin{figure}[h]
	\centering
	\includegraphics[height=0.9\textheight]{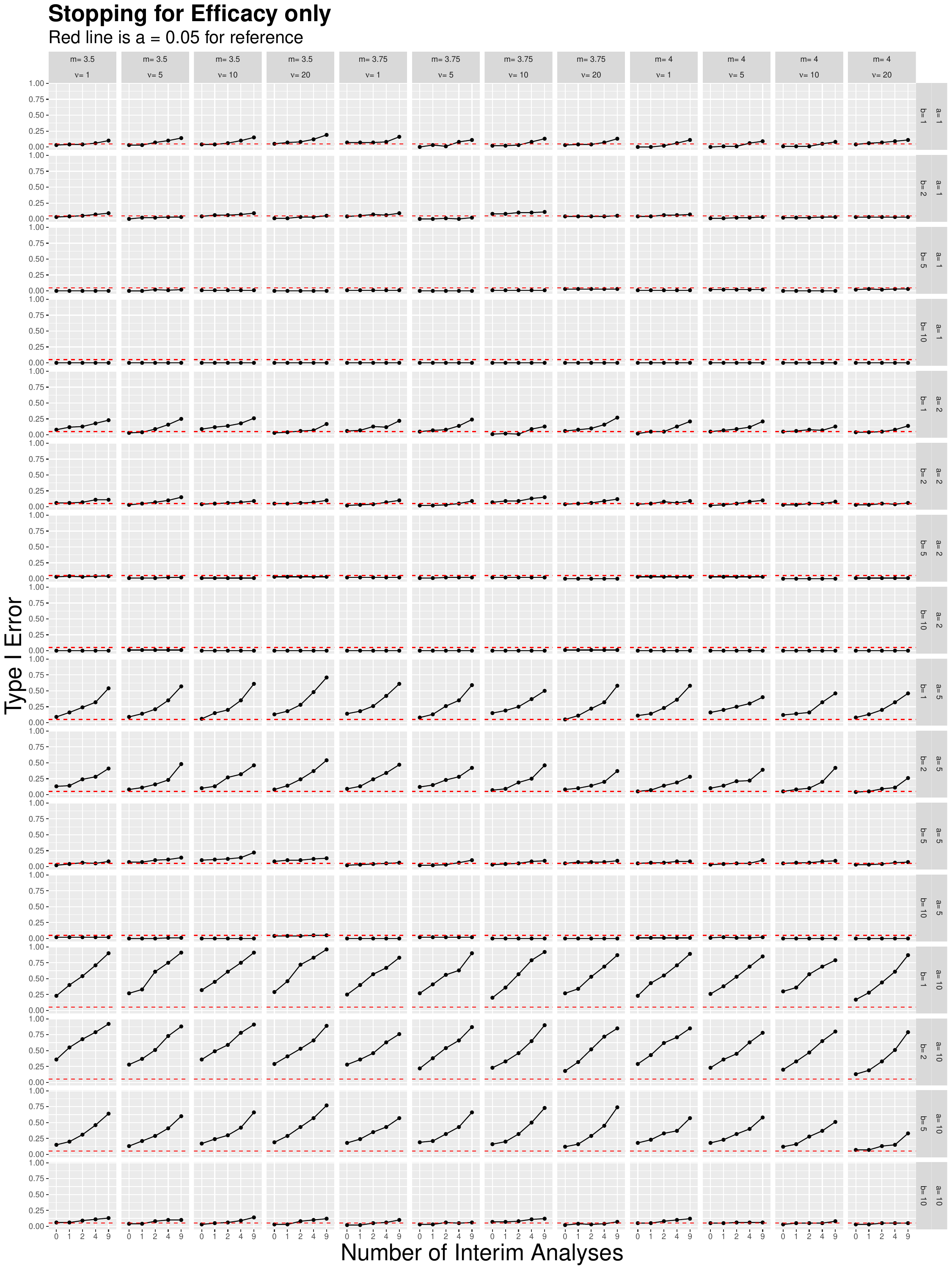}
	\caption{Type I errors when stopping for efficacy for simulation study prior sensitivity analysis.}
	\label{fig:sm-efficacy-type1}
\end{figure}

Figures \ref{fig:sm-either-type1}, \ref{fig:sm-futility-type1}, \ref{fig:sm-efficacy-type1} show Type I error rates for stopping for either, stopping for futility only, and stopping for efficacy only, respectively. The model makes an unacceptable number of type I errors when the ratio $b/a > 2$, and does see degradation of type I error at a high number of interim analyses for $a/b=2$. This holds for all three stopping scenarios. This is not surprising since $a$ represents the number of prior ``success'' and $b$ the number of prior ``failures'', so a model using this prior has a strong prior on the measure being met, when in fact it is not. Looking back at  Figure \ref{fig:priors}, these priors correspond to scenarios where most have $P(\phi > 0.8) > 0.8$, and approaching 1 in some cases. When stopping for futility only, the type I errors don't change with number of interim analyses like in the either or efficacy cases. Although there are small differences due to changes in $m$ and $\nu$, they have a smaller effect than $a$ and $b$.

\begin{figure}[h]
	\centering
	\includegraphics[height=0.9\textheight]{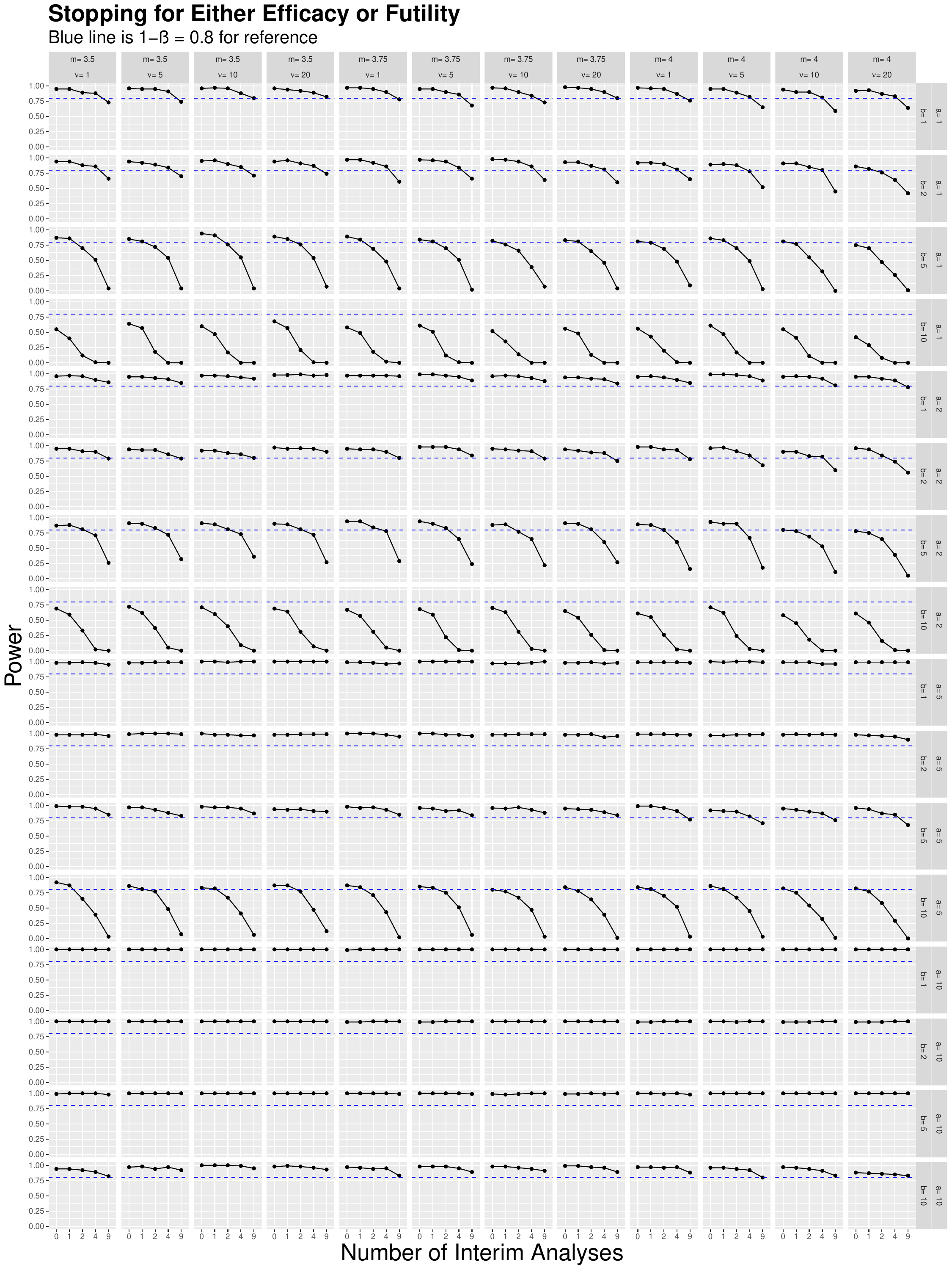}
	\caption{Power when stopping for either futility or efficacy for simulation study prior sensitivity analysis.}
	\label{fig:sm-either-power}
\end{figure}

\begin{figure}[h]
	\centering
	\includegraphics[height=0.9\textheight]{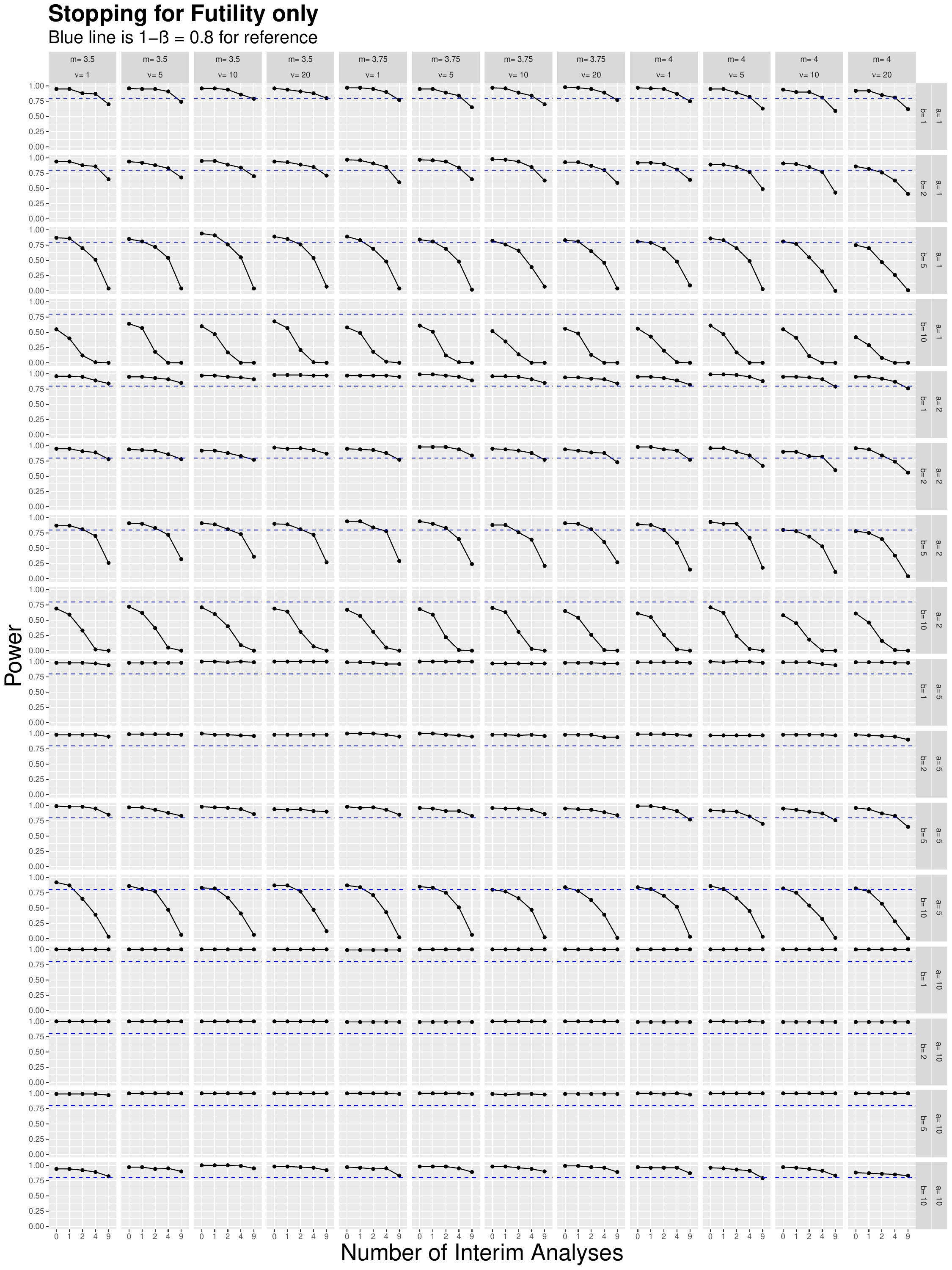}
	\caption{Power when stopping for  futility  for simulation study prior sensitivity analysis.}
	\label{fig:sm-futility-power}
\end{figure}

\begin{figure}[h]
	\centering
	\includegraphics[height=0.9\textheight]{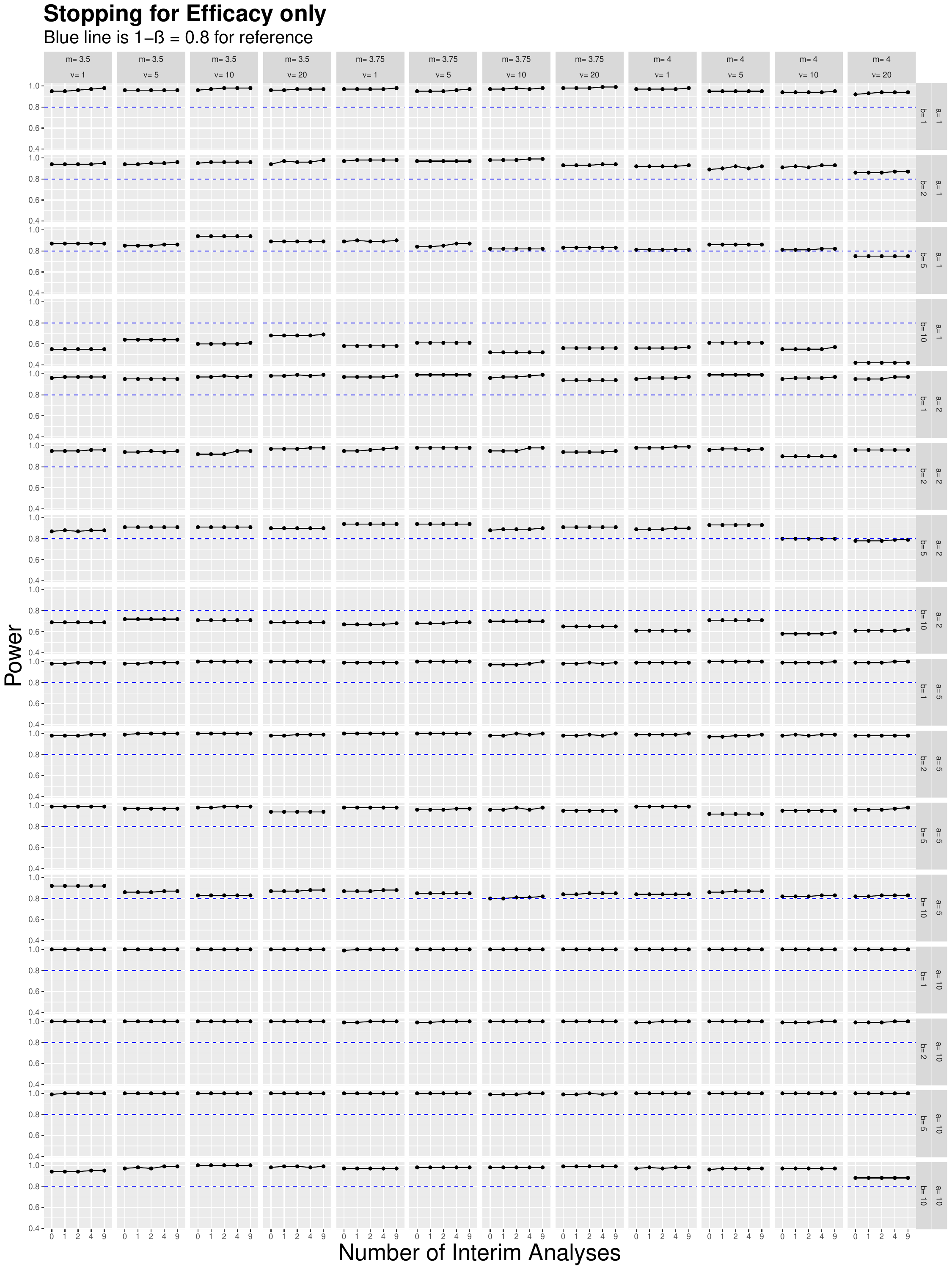}
	\caption{Power when stopping for  efficacy for simulation study prior sensitivity analysis.}
	\label{fig:sm-efficacy-power}
\end{figure}

Figures \ref{fig:sm-either-power}, \ref{fig:sm-futility-power}, \ref{fig:sm-efficacy-power} show power rates for stopping for either, stopping for futility only, and stopping for efficacy only, respectively. Stopping for either and stopping for futility have similar results, with power dropping below 0.8 when $b/a > 2$, particularly at a higher number of interim analyses. When only stopping for futility, power is above 0.8 except when $b/a > 5$. 

\begin{figure}[h]
	\centering
	\includegraphics[height=0.9\textheight]{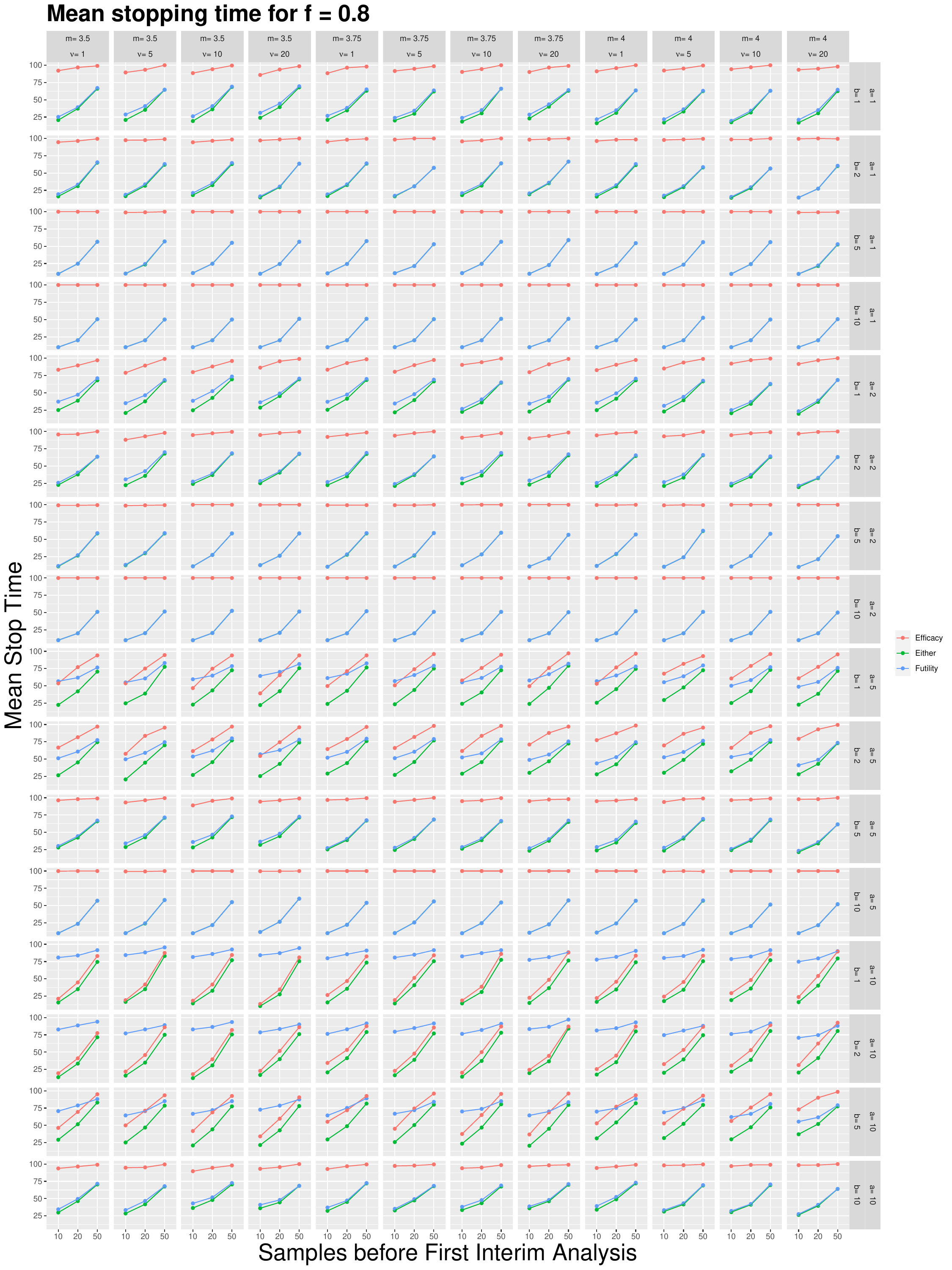}
	\caption{Mean stopping time when $\phi=0.8$ (i.e. measure is not met) for simulation study prior sensitivity analysis.}
	\label{fig:sm-mean-stopping-08}
\end{figure}

\begin{figure}[h]
	\centering
	\includegraphics[height=0.9\textheight]{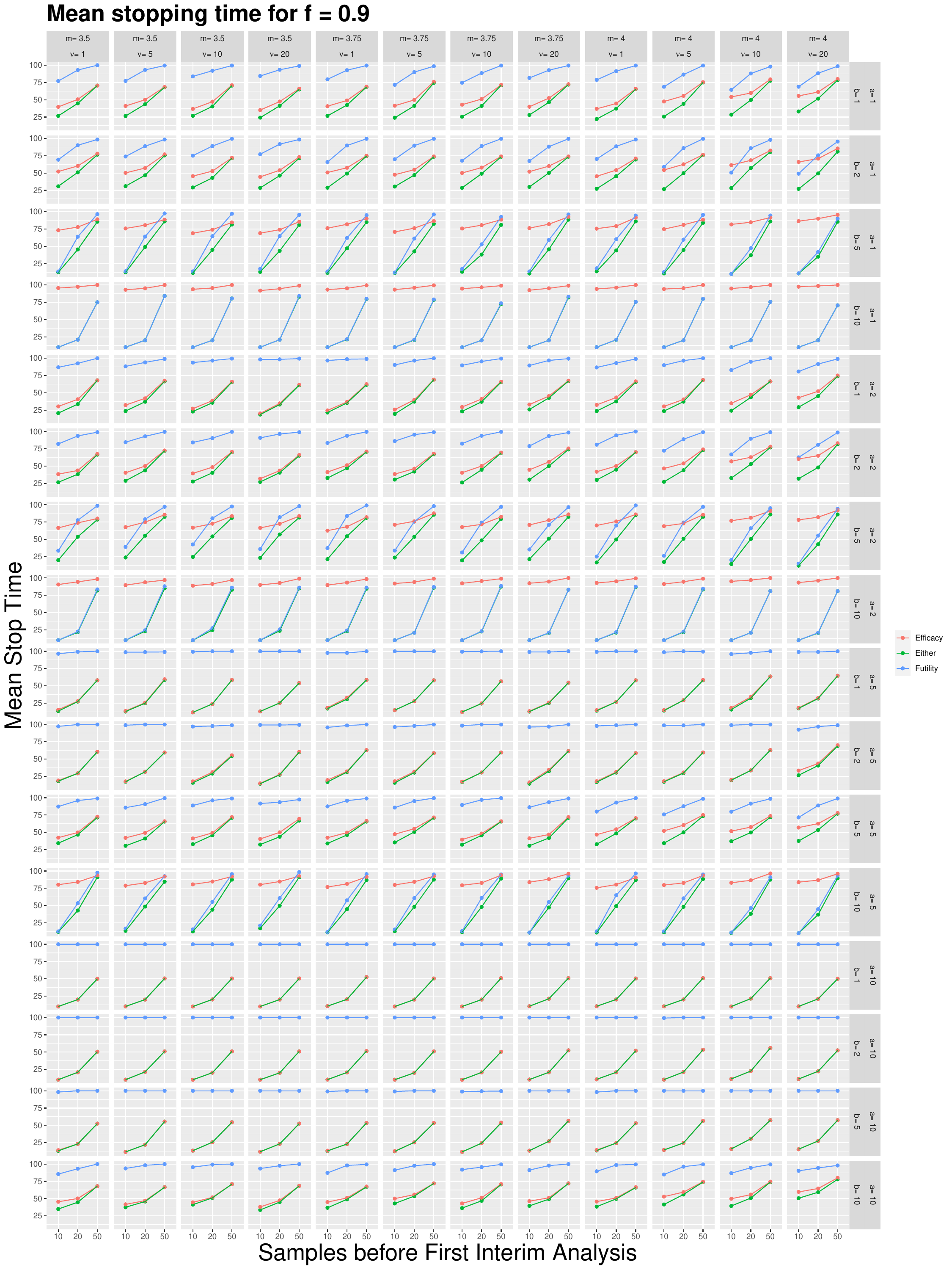}
	\caption{Mean stopping time when $\phi=0.9$ (i.e. measure is  met) for simulation study prior sensitivity analysis.}
	\label{fig:sm-mean-stopping-09}
\end{figure}

Figure \ref{fig:sm-mean-stopping-08} and \ref{fig:sm-mean-stopping-09} show mean stopping times for $\phi=0.8$ and $\phi=0.9$, respectively. For $\phi=0.8$, the truth is the measure is not met, which is why stopping for futility only tends to stop the earliest, except in cases where $a/b > 2$, which is where type I errors are occurring due to stopping early and claiming the measure is met. When the ratio $a/b = 1$, stopping for futility (or either) often occurs slightly after (10-20 pulls extra) the first interim analysis, on average. For $\phi=0.9$, the truth is the measure is  met, which is why stopping for efficacy only tends to stop the earliest, except in cases where $b/a > 2$, which is where the prior is very pessimistic and type II errors are commonly made. When the ratio $a/b = 1$, stopping for efficacy (or either) often occurs slightly after (10-20 pulls extra) the first interim analysis, on average.

\subsection*{Prior Sensitivity Analysis for Binary Model Case}		

In order to understand the influence of the selected prior of Beta$(0.6, 0.4)$ on the results, a sensitivity analysis was conducted using the following priors:
\begin{itemize}
	\item Prior 1: Beta$(0, 0)$ (non-informative improper prior)
	\item Prior 2: Beta$(1,1)$ (a flat prior, used above)
	\item Prior 3: Beta$(6, 4)$ (a more informative prior)
\end{itemize}

\begin{figure}[h]
	\centering
	\includegraphics[width=0.7\linewidth]{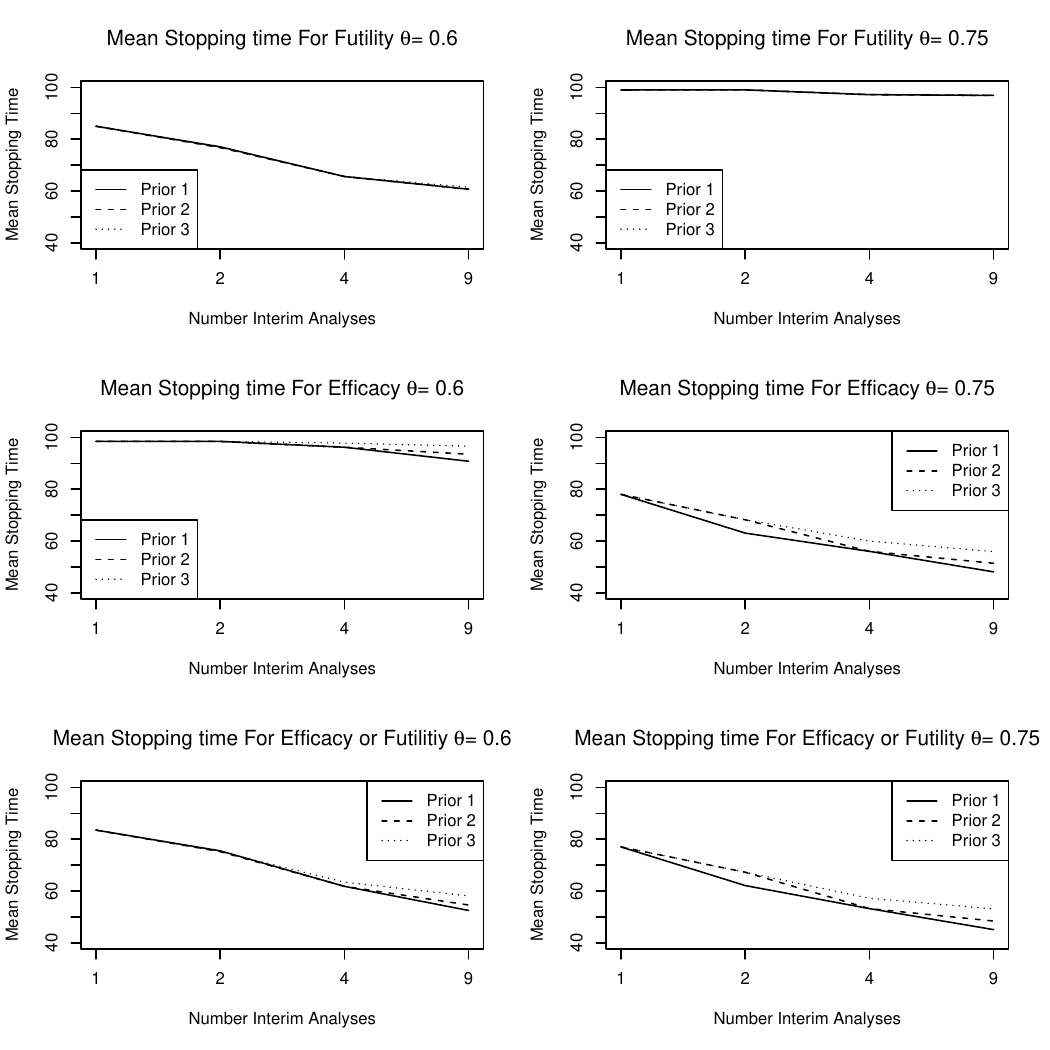}
	\caption{Average stopping time  for Binomial case simulation study prior sensitivity analysis.}
	\label{fig:mean-stoppingsensitivity-analysis}
\end{figure}

Figure \ref{fig:mean-stoppingsensitivity-analysis} explores the stopping behavior, on average, of a test that is stopped for efficacy and/or futility, based on the number of samples seen before the first interim analysis was conducted for each of the selected priors for sensitivity analysis. We find the average stopping rates to be similar across all three priors, which the reference prior generally stopping slightly earlier than when using the other priors.

\begin{figure}[h]
	\centering
	\includegraphics[width=0.7\linewidth]{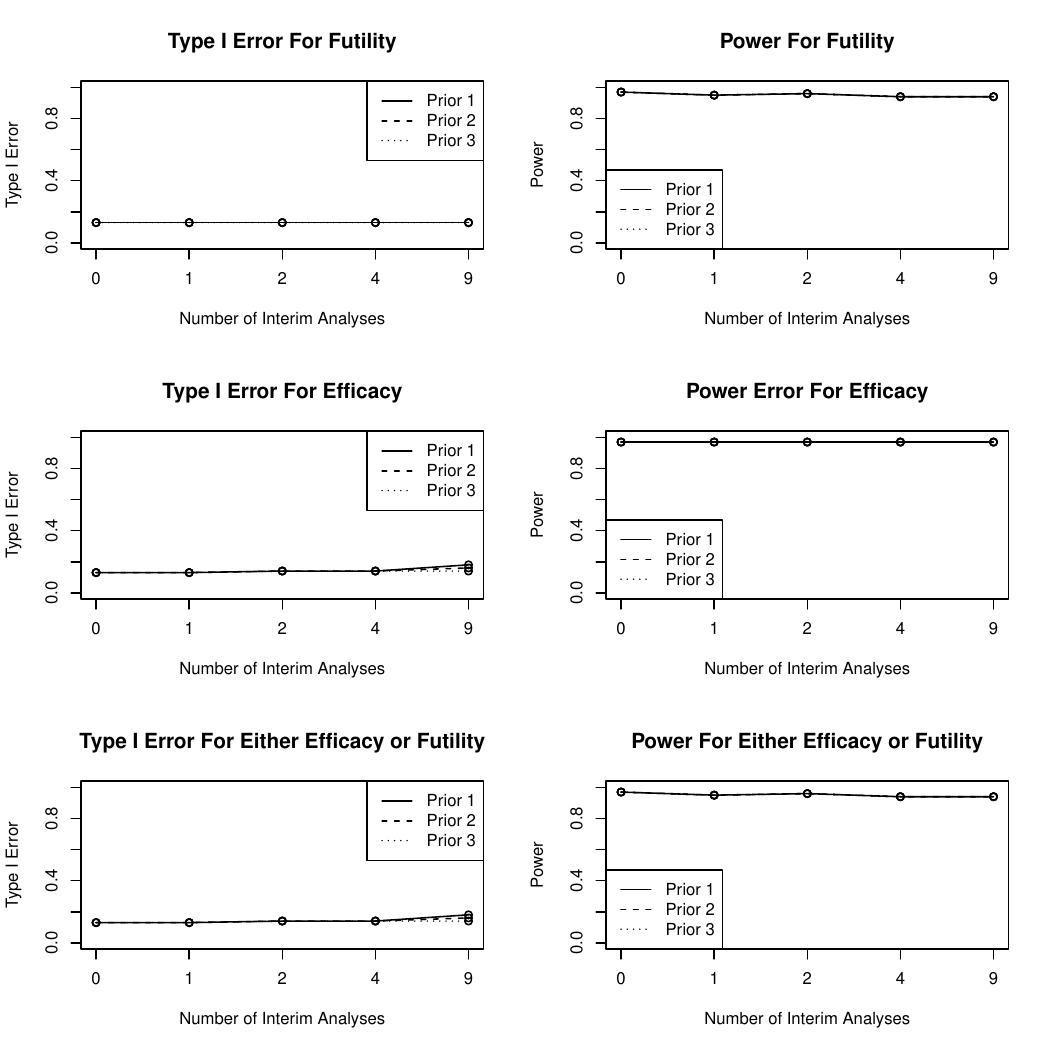}
	\caption{Type I error and power  for Binomial case simulation study prior sensitivity analysis.}
	\label{fig:sensitivity-plots}
\end{figure}

To understand the operating characteristics of these design constructs under different priors, Figure \ref{fig:sensitivity-plots} demonstrates type I error and power for PP when the test can stop early for efficacy only, futility only, and either efficacy or futility based on different number of interim analyses. While power remains unaffected by the prior choice in this example, type I error rate is worse for the reference prior---likely due to the earlier stopping times.

\subsection*{Prior Sensitivity Analysis for Application}

\begin{figure}[h]
	\centering
	\includegraphics[width=\textwidth]{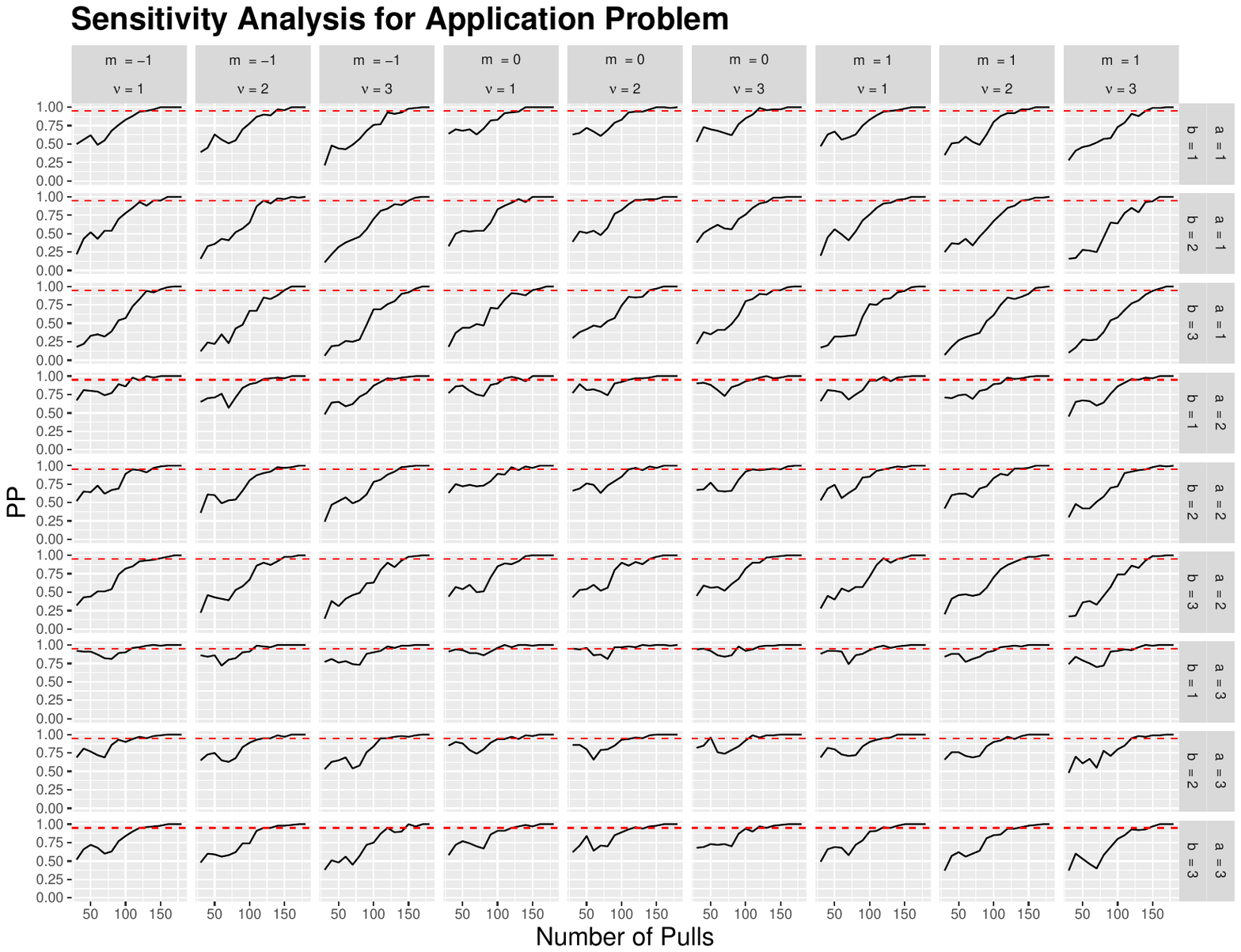}
	\caption{Prior sensitivity analysis for application in main paper.}
	\label{fig:sm-sens-analysis}
\end{figure}

Figure \ref{fig:sm-sens-analysis} shows a sensitivity analysis for the application problem. Here, the metric on the y-axis is PP directly.  In all cases considered, models would recommend early stopping and claiming the measure was met, the differences being the number of pulls when this happened. This shows some degree of robustness, at least for the range of priors considered. The most informative priors have a ratio $b/a = 3$, with less influence by $m$ and $\nu$, as seen in the sensitivity analysis of the simulation study.

\end{document}